\documentclass[11pt]{article}

\usepackage{amsmath,amssymb}
\usepackage{graphicx}
\usepackage{amscd}
\usepackage{theorem}

\newcommand{\R}{{\mathbb{R}}}
\newcommand{\Z}{{\mathbb{Z}}}
\newcommand{\C}{{\mathbb{C}}}
\newcommand{\T}{{\mathbb{T}}}

\newcommand{\ba}{\begin{array}}
\newcommand{\ea}{\end{array}}

\newcommand{\bp}{\begin{pmatrix}}
\newcommand{\ep}{\end{pmatrix}}

\newcommand{\bps}{\begin{smallmatrix}}
\newcommand{\eps}{\end{smallmatrix}}

\newcommand{\ti}{\tilde}

\newcommand{\la}{\langle}
\newcommand{\ra}{\rangle}

\def \({\left(}
\def \){\right)}

\def \cA{{\cal A}}

\def \cC{{\cal C}}

\def \cF{{\cal F}}
\def \cH{{\cal H}}

\def \cS{{\cal S}}

\def \cZ{{\cal Z}}
\def \V{{ \cal V}}

\DeclareMathOperator{\Tr}{Tr}

\def \qed{\hfill $\blacksquare$}

\def \lraw{\leftrightarrow}
\def \raw{\rightarrow}

\def \hraw{\hookrightarrow}

\def \Re{\mathrm{Re}}
\def \Im{\mathrm{Im}}

\def \deg{\mathrm{deg}}
\def \rank{\mathrm{rank}}

\def \dim{\mathrm{dim}}
\def \Ed#1#2{\mathrm{End}_{#1} {#2}}
\def \Hom{\mathrm{Hom}}
\def \Ob{\mathrm{Ob}}

\def \Arg{\mathrm{Arg}}

\def \half{\frac{1}{2}}
\def \ov#1{\frac{1}{#1}}
\def \zb{\bar{z}}

\def \Eh{\hat{E}}

\def \delm#1{{\delta_{[#1]}}}

\def \fpartial#1{\frac{\partial}{\partial {#1}}}

\def \wt#1{{\widetilde {#1}}}

\def \f{{\frak f}}

\def \l{{\frak l}}
\def \m{{\frak m}}
\def \n{{\frak n}}

\def \l({\left(}
\def \r){\right)}

\def \0{{\bf 0}}
\def \1{{\bf 1}}

\def \ie{{\it i.e.\ }}

\def \gh{{\hat g}}

\def \Mh{{\hat M}}
\def \Th{{\hat \T}}
\def \tauh{{\hat \tau}}
\def \rhoh{{\hat \rho}}

\def \nabb{{\bar \nabla}}
 
\def \lb{{\bar l}}

{\theorembodyfont{\rmfamily}
 \newtheorem{defn}{Definition}[section]
 \newtheorem{thm}{Theorem}[section]
 \newtheorem{lem}{Lemma}[section]
 
 \newtheorem{prop}{Proposition}[section]
 \newtheorem{cor}{Corollary}[section]
 \newtheorem{rem}{Remark}[section]
 \newtheorem{conj}{Conjecture}[section]
}

\def \pf{{\it proof.}\quad}

\begin{document}

\begin{titlepage}
\thispagestyle{empty}
\begin{flushleft}
\hfill YITP-04-34 \\
\hfill hep-th/0406233\\
\hfill Jun, 2004 \\
%\today
\end{flushleft}

\vskip 1.5 cm

\begin{center}
\noindent{\Large \textbf{Homological mirror symmetry 
on noncommutative two-tori  }}\\

\noindent{
 }\\
\renewcommand{\thefootnote}{\fnsymbol{footnote}}

\vskip 2cm
{\large 
Hiroshige Kajiura 
\footnote{e-mail address: kajiura@yukawa.kyoto-u.ac.jp}\\

\noindent{ \bigskip }\\

\it
Yukawa Institute for Theoretical Physics, Kyoto University \\
Kyoto 606-8502, Japan\\
}

\bigskip
\end{center}
\begin{abstract}

Homological mirror symmetry is a conjecture that 
a category constructed in the A-model and a category constructed 
in the B-model are equivalent in some sense. 
We construct a cyclic differential graded (DG) category of 
holomorphic vector bundles on noncommutative two-tori 
as a category in the B-model side. 
We define the corresponding Fukaya's category in the A-model side, and 
prove the equivalence of the two categories at the level of cyclic categories. 
We further write down explicitly Feynman rules for higher Massey products 
derived from the cyclic DG category. 
As a background of these arguments, a physical explanation of the 
mirror symmetry for noncommutative two-tori is also given.

\end{abstract}
\vfill

\end{titlepage}
\vfill
\setcounter{footnote}{0}
\renewcommand{\thefootnote}{\arabic{footnote}}
\newpage

\tableofcontents

\section{Introduction}

String theories have provided us with various `stringy' 
deformations of the geometries of their target spaces. 
In general closed strings include gravitons, which induce the
deformation of the metric on the target space. 
Open strings include gauge fields, which define the connections 
on vector bundles (D-branes). 
Thus, closed string physics is related directly to the geometry of the 
target space $M$, whereas open string physics in general extracts 
some informations of the geometry in a more abstract algebraic way. 
Various string dualities are then interpreted as some equivalences between 
the free energy, \ie, a collection of correlation 
functions, of string theories. 
One of such string dualities is mirror symmetry, 
a symmetry between Calabi-Yau manifolds, 
which can be interpreted by topological closed string physics. 
There are two types of topological strings 
whose target spaces are Calabi-Yau manifolds. 
One is called the A-model, which depends on the complexified 
symplectic structure and is independent of the complex structure 
of the Calabi-Yau manifold. 
Another one, the B-model in contrast depends on the complex
structure only. 
For a given Calabi-Yau manifold $M$, the mirror symmetry conjecture 
claims the existence of a mirror Calabi-Yau manifold $\Mh$ 
such that the A-model closed string on $M$ is equivalent to 
the B-model closed string on $M$ and vice versa \cite{W1}. 
Homological mirror symmetry conjecture proposed by 
Kontsevich \cite{mirror} is 
thought of as an open string version \cite{W2} 
of this mirror symmetry conjecture. 
Open string theory in general includes some kind of D-branes. 
It forms a ``D-brane category'' (see \cite{Dbcat}); 
D-branes and open strings are identified with objects and morphisms 
between the objects, respectively, and the free energy determines the composition 
rules of the morphisms. 
For the tree open string A-model, the corresponding category is 
Fukaya's $A_\infty$-category \cite{Fukaya}. 
On the other hand, what is constructed on the B-model side is 
a category of holomorphic vector bundles or coherent sheaves 
more generally. 
The homological mirror symmetry conjecture then states that 
the Fukaya category on a Calabi-Yau manifold $M$ is in some sense 
equivalent to the category of coherent sheaves on the mirror dual 
Calabi-Yau manifold $\Mh$. 
Now, the conjecture is checked successfully 
in the case $M$ is an elliptic curve \cite{mirror,PZ,Poli1,Poli2}, 
an abelian variety \cite{F,KoSo}, a quartic surface \cite{Sei}, 
and so on. 

Since open string physics describe the target space geometry algebraically, 
it allows us to consider more extended geometry than 
the classical ones just as noncommutative geometry \cite{co-book} does. 
Some sort of noncommutative geometry are in fact interpreted by 
open string theories \cite{CDS,DH,KO,CK,Scho,SW,CF}. 
In the spirit of noncommutative geometry, 
a noncommutative algebra is regarded as the space of functions on a 
`noncommutative space'. 
Projective modules over the noncommutative algebra are then 
vector bundles, \ie D-branes over the noncommutative space. 
Thus, one can consider a D-brane category 
over the noncommutative space. 
K-theory is then defined in the framework of $C^*$-algebra, 
where the $K_0$-group consists of the projective modules. 
Since the Bott periodicity holds as in the case of 
usual commutative spaces, one can say that 
a noncommutative $C^*$-algebra describes some kind of space. 
Here, although the corresponding open string theory is different from 
the topological strings in the mirror symmetry set-up, 
if for instance 
a complex structure would be defined on some kind of noncommutative
Calabi-Yau manifold, 
one may consider a category of B-model on it. 

For such a direction, an ideal candidate is that given by 
A.~Schwarz \cite{Stheta,Stensor}, where a complex structure is
introduced on even dimensional noncommutative tori and 
a holomorphic structure is defined on projective modules over it. 
In fact for noncommutative (real) two-tori, 
we succeeded to find the corresponding A-model side and 
observe a part of the homological mirror symmetry in the previous paper 
\cite{foliation}. 
For a recent discussion see also \cite{KimKim}. 
The criterion to find the A-model side \cite{foliation} 
is the T-duality in string theory. 
Now, noncommutative tori are interpreted from the physics of 
open strings \cite{CDS,DH,KO,CK,SW}, and in particular in \cite{foliation} 
we explained that two different representations of noncommutative two-tori 
are related by T-duality \cite{foliation}. 
It is known that the T-duality coincides with the mirror symmetry 
for two-tori \cite{DVV}. Introducing a complex and holomorphic 
structure on one side and transforming it by the T-duality 
we can expect what kind of category should be taken for the A-model side \cite{foliation}. 
In this two-dimensional case, the A-model side is essentially
independent of the noncommutativity $\theta\in\R$ 
and the same as the commutative case in \cite{PZ}. 
This means the B-model side, the category of 
holomorphic vector bundles over a noncommutative two-torus, 
should be essentially independent of $\theta$. 
Such thought leads to show that the derived categories of 
holomorphic vector bundles over noncommutative two-tori 
with different $\theta$ are equivalent \cite{PoSc}. 
There a relation between the noncommutativity $\theta$ and 
the stability condition introduced by Douglas, Bridgeland \cite{Dou,Bridge} 
is also discussed.

This paper is a continuation of our previous work \cite{foliation}. 
We construct a cyclic differential graded (DG) category of 
holomorphic vector bundles on noncommutative two-tori as a category of the B-model side. 
We clarify the meaning of the mirror symmetry for noncommutative two-tori, 
especially the relation of the noncommutativity $\theta$ with the $t$-structure, 
and set up the homological mirror symmetry on noncommutative two-tori. 
As a part of it, we prove 
the equivalence of the cyclic DG category to 
the corresponding category in A-model side as cyclic categories. 
We further write down explicitly Feynman rules for higher Massey $A_\infty$-products 
derived from the cyclic DG category. 

One may take the zero noncommutativity limit in these results. 
They then reproduce various results in the commutative case 
(see \cite{mirror,PZ,Poli1,Poli2}) in a different description. 
One of the advantage of employing these noncommutative tools there is 
the explicitness. 
They provide us with a manifestly $SL(2,\Z)$ covariant description, where 
$SL(2,\Z)$ is the symmetry of noncommutative tori known as the Morita equivalence 
(see \cite{Rtwo,co-book,KS}), and further 
morphisms of the cyclic DG category can be described explicitly 
by Hermite polynomials in general. 
An alternate feature of our approach is that 
we treat carefully the cyclic structure which the categories have.

In section \ref{sec:nctori} we reformulate some tools for 
noncommutative two-tori and differential geometry on it. 
In subsection \ref{ssec:nctori}, we recall the definitions of 
noncommutative two-tori and Heisenberg modules, which are 
projective modules over noncommutative tori equipped with 
constant curvature connections. 
The space of homomorphisms between two Heisenberg modules is also identified with 
a Heisenberg module. As the composition of the homomorphisms, 
the tensor product between two Heisenberg modules are given in 
subsection \ref{ssec:tensor}. 
In subsection \ref{ssec:string} we explain the underlying physics of open string 
theory. 
{}From the viewpoint of {\it mirror symmetry for noncommutative two-tori}, 
the meaning of the noncommutativity is clarified. 
Although the reader can skip this subsection mathematically, it gives a criterion to set up 
homological mirror symmetry conjecture on noncommutative two-tori. 
In section \ref{sec:ncmirror} we discuss a complex geometry of noncommutative two-tori, 
its dual symplectic side, and the correspondence between them, the homological mirror symmetry. 
In subsection \ref{ssec:hol} we recall the holomorphic structures on 
Heisenberg modules introduced by A.~Schwarz \cite{Stheta,Stensor} 
and construct a cyclic differential graded (DG) category of 
(stable) holomorphic vector bundles over a noncommutative two-torus 
partially based on the work by Polishchuk-Schwarz\cite{PoSc}. 
In subsection \ref{ssec:cat} we propose a candidate of the corresponding category 
on the A-model side, 
the Fukaya's $A_\infty$-category for noncommutative two-tori, 
and set up a homological mirror symmetry conjecture on 
noncommutative two-tori. 
As a part of the conjecture, we prove a categorical mirror symmetry 
in a stronger sense than the Polishchuk-Zaslow's result 
for (commutative) elliptic curve \cite{PZ}. 
Finally in subsection \ref{ssec:main} 
we complete the Feynman rule to generate higher Massey $A_\infty$-products, 
which form a minimal cyclic $A_\infty$-category, from the cyclic 
DG category of holomorphic vector bundles on noncommutative two-tori. 
In this paper, we treat any (graded) vector space over field $k=\C$.

 \section{Noncommutative two-tori}
\label{sec:nctori}

In this section we recall and arrange some basic tools for noncommutative two-tori 
to apply them to a mirror symmetry set-up. 
The definitions of noncommutative two-tori and Heisenberg modules are recalled 
in subsection \ref{ssec:nctori}, 
and the tensor products between two Heisenberg modules are given in 
subsection \ref{ssec:tensor}. 
In subsection \ref{ssec:string} we explain the underlying open string
theory physics, which gives a criterion to set up 
homological mirror symmetry conjecture on noncommutative two-tori. 
In this paper, we define noncommutative tori and 
various structures on them in a different notations from a conventional one 
in order for them to fit the notation of the mirror symmetry set-up.

 \subsection{Noncommutative two-tori and Heisenberg modules}
\label{ssec:nctori}

A noncommutative two-torus $\cA_\theta$ is defined by two unitary generators 
$U_1, U_2$ with relation 
\begin{equation}
 U_1 U_2=e^{-2\pi i\theta}U_2 U_1\ .
 \label{theta-rel}
\end{equation}
Any element $a\in\cA_\theta$ is represented as 
\begin{equation*}
 a=\sum_{(n_1,n_2)\in\Z^2}a_{n_1n_2}(U_1)^{n_1}(U_2)^{n_2}\ ,
 \quad a_{n_1n_2}\in\C\ ,
\end{equation*}
where $a$ belongs to the Schwartz space $\cS(\Z^2)$. 
The trace is defined by 
\begin{equation*}
 \Tr(a)=a_{00}\ , 
\end{equation*}
which is just the integral on $\T^2$ when $\theta=0$. 
In this paper, we let $\theta\in\R$ irrational for simplicity. 
But the results of this paper reduce to the ordinary commutative
two-tori case by taking the limit $\theta\raw 0$, 
and the rational $\theta$ cases are in fact equivalent to the $\theta=0$
case. 
A representation of this noncommutative two-torus is the 
rotation algebras (see \cite{co-book}). 
Consider the space of complex-valued continuous functions $C(S^1)$ on $S^1$ 
parametrized by $x\in\R$ with periodicity $x\sim x+1$. 
$U_1,U_2:C(S^1)\raw C(S^1)$ are defined by 
\begin{equation*}
 U_1 a(x)=e^{2\pi i x} a(x)\ ,\qquad  U_2 a(x)=a(x+\theta)\ .
\end{equation*}
The two generators $U_1$, $U_2$ generates the crossed product algebra, 
which is a noncommutative two-torus $\cA_\theta$ above.

In noncommutative geometry, the analogue of vector bundles are 
projective modules. 
On noncommutative tori, 
projective modules, called Heisenberg modules, 
are given explicitly (see \cite{KS}). 
For each $g=(\bps q & s\\p & r\eps)\in SL(2,\Z)$, 
if $p=0$ we just take $C(S^1)$ above as a module. If $p\ne 0$, 
a Heisenberg module $E_{g,\theta}$ over $\cA_\theta$ 
is given by the space of $|p|$ copies of functions on $\R$, or more precisely, 
the Schwartz space $\cS(\R\times\Z_{|p|})$. 
On $f\in E_{g,\theta}$, the action of $\cA_\theta$ is defined
by 
\begin{equation}
 \begin{split}
 (U_1 f)(x,j)&=f(x,j)e^{2\pi i(x+j\frac{q}{p})}\\
 (U_2 f)(x,j)&=f(x+\frac{q}{p}+\theta,j-1)\ ,
 \end{split} \label{H-mod}
\end{equation}
where $x\in\R$ and $j=0,\cdots,p-1$. 
One can see that $U_1$ and $U_2$ in fact satisfy 
the noncommutativity relation (\ref{theta-rel}). 
In general, for a projective module $E$ over a $C^*$-algebra $\cA$, 
the space of endomorphisms of $E$ that commute with the action of $\cA$, 
denoted by $\Ed{\cA}E$, forms an algebra, and it is called 
Morita equivalent to $\cA$. 
Such a projective module $E$ is then called a Morita equivalence 
bimodule. In the case of noncommutative tori, 
the endomorphism algebra $\Ed{\cA_\theta}E_{g,\theta}$ is also 
generated by two generators $Z_1,Z_2$. 
It is given as right actions by 
\begin{equation}
 \begin{split}
 (f Z_1)(x,j)&=f(x,j)e^{2\pi i(\frac{x}{q+p\theta}+\frac{j}{p})}\\
 (f Z_2)(x,j)&=f(x+\ov{p},j-r)\ .
 \label{Ed}
 \end{split}
\end{equation}
These generators satisfy the following relation
\begin{equation}
 Z_1Z_2=e^{-2\pi i(g\theta)}Z_2Z_1\ , 
 \label{Z-rel}
\end{equation}
where $g\theta$ is defined by 
\begin{equation}
 g\theta=\frac{r\theta+s}{q+p\theta}\ . \label{g-theta}
\end{equation}
Thus, the endomorphism algebra $\Ed{\cA_\theta}E_{g,\theta}$ generated by $Z_1$ and $Z_2$ again 
forms a noncommutative torus $\cA_{g\theta}$, 
and $E_{g,\theta}$ is a Morita equivalence $\cA_\theta$-$\cA_{g\theta}$ bimodule. 
It is used to show explicitly that $\cA_\theta$ and $\cA_{g\theta}$ 
are Morita equivalent (see\cite{Rhigh,RS,KS}). 
It is known that any finitely generated projective module 
over $\cA_\theta$ is isomorphic to the direct sum of Heisenberg
modules $E_{g,\theta}$ with $g\in SL(2,\Z)$
(see\cite{CR,KS}). 
In such a sense $E_{g,\theta}$ for each $g\in SL(2,\Z)$ is called 
a {\it basic module} in \cite{PoSc}.  
This fact implies that it is enough to concentrate our arguments to these basic modules. 

A connection $\nabla_i: E_{g,\theta}\raw E_{g,\theta}$, $i=1,2$ is
defined so that 
\begin{equation*}
 \nabla_i(U_j f)= (2\pi i\delta_{ij}U_j) f+ U_j (\nabla_i f)
\end{equation*}
holds. Here the first term of the right hand side means that 
$\nabla_i$ acts to $U_j$ as a derivation for $i$-direction. 
It is known that $E_{g,\theta}$ can be equipped the following 
constant curvature connection. 
\begin{equation}
 \nabla_1=\fpartial{x}-\frac{2\pi i\beta}{q+p\theta}\ ,\quad 
 \nabla_2=-\frac{2\pi i p}{q+p\theta}x-\frac{2\pi i\alpha}{q+p\theta}\ ,
 \qquad \alpha,\beta\in\R\ .
 \label{ccc}
\end{equation}
Thus, modules $E_{g,\theta}$ has a constant curvature 
$[\nabla_1,\nabla_2]=-\frac{2\pi i p}{q+p\theta}$. 
In eq.(\ref{ccc}), $\alpha$ and $\beta$ parametrize the moduli of 
constant curvature connection \cite{CR,KS}. By gauge transformation 
$\nabla_i \raw (Z_j)^{-1}\nabla_i Z_j$, we have 
$\alpha\sim \alpha+1$ and 
$\beta\sim \beta+1$.
\footnote{Here we change the notation of $\alpha_a$ and $\beta_a$ 
from that in \cite{foliation}. $q_a\alpha_a$ and $q_a\beta_a$ 
in this paper coincide with $-\alpha_a$ and $\beta_a$ 
in \cite{foliation}, respectively. }

Next, we would like to consider the space of homomorphisms 
$\Hom(E_{g_a,\theta},E_{g_b,\theta})$ 
between two projective modules $E_{g_a,\theta}$ and $E_{g_b,\theta}$, 
where $g_a,g_b\in SL(2,\Z)$. In fact, 
$\Hom(E_{g_a,\theta},E_{g_b,\theta})$ can be identified with 
an $\cA_{\theta_a}$-$\cA_{\theta_b}$ bimodule, where $\theta_a:=g_a\theta$. 
If $g_a=g_b$, it is just $\cA_{\theta_a}$, whereas 
if $g_a\ne g_b$, it is the Heisenberg module 
$E_{g_{ab},\theta_a}$, where we introduced 
the following notation 
\begin{equation}
 g_{ab}=g_a^{-1}g_b
 =\bp r_a & -s_a\\ -p_a & q_a\ep
             \bp q_b & s_b\\  p_b & r_b\ep
 =\bp q_{ab} &  s_{ab}\\  p_{ab} & r_{ab}\ep\ .
 \label{notation}
\end{equation}
There exists a natural identification between 
$\cA_{\theta_a}$-$\cA_{\theta_b}$ bimodule and 
$\cA_{\theta_b}$-$\cA_{\theta_a}$ bimodule. 
Namely, one can construct a canonical map $E_{g_{ab},\theta_a}\raw E_{g_{ba},\theta_b}$, 
which we denote by $\dag$. If $p_{ab}=0$, 
for $a=a_{n_1n_2}(U_1)^{n_1}(U_2)^{n_2}\in E_{g_{ab},\theta_a}=\cA_{\theta_a}$ 
$a^\dag$ is simply 
\begin{equation*}
 a^\dag=(a_{n_1n_2})^*(U_2)^{-n_2}(U_1)^{-n_1}\quad\in \cA_{\theta_a}\ , 
\end{equation*}
where $*$ is the complex conjugate. 
For $p_{ab}\ne 0$, the relation between 
$f_{ab}\in E_{g_{ab},\theta_a}$ and 
$f_{ab}^\dag\in E_{g_{ba},\theta_b}$ is given by 
\begin{equation}
 f_{ab}(x,j)= 
 (f_{ab}^\dag)^*\(\frac{x}{q_{ab}+p_{ab}\theta_a},-j q_{ab}\)\ .
 \label{dag}
\end{equation}  
\begin{lem}
For $f_{ab}\in E_{g_{ab},\theta_a}$, $p_{ab}\ne 0$ and 
$f_{ab}^\dag\in E_{g_{ba},\theta_b}$ defined in eq.(\ref{dag}), 
\begin{equation*}
 \begin{split}
  &(U_i (f_{ab}^\dag))^*\(\frac{x}{q_{ab}+p_{ab}\theta_a},-j q_{ab}\)
  =(f_{ab} Z_i)(x,j)\ ,\\
  &((f_{ab}^\dag) Z_i)^*\(\frac{x}{q_{ab}+p_{ab}\theta_a},-j q_{ab}\)
  =(U_i f_{ab})(x,j)\ 
 \end{split}
\end{equation*}
hold. Here note that $U_i$ and $Z_i$ in the left hand side act on 
$E_{g_{ba},\theta_b}$, whereas, those in the right hand side 
act on $E_{g_{ab},\theta_a}$. 
 \label{lem:dag}
\end{lem}
It can be checked by a direct calculation. 
This lemma implies how the bijection $\dag$ has good properties.
\begin{defn}[$\Tr$ on $\cA_{\theta_a}$]
For $a=\sum_{(n_1,n_2)\in\Z^2}a_{n_1n_2}(U_1)^{n_1}(U_2)^{n_2}$, 
$\Tr_{\cA_{\theta_a}}: \cA_{\theta_a}\raw\C$ is defined by 
\begin{equation}
 \Tr_{\cA_{\theta_a}} a :=|q_a+p_a\theta|a_{00}\ .
\end{equation}
 \label{defn:trace}
\end{defn}
Using this trace, the topological charge of noncommutative 
vector bundle $E_{g_a,\theta}$ over $\cA_\theta$ is defined by 
$\Tr_{\cA_{\theta_a}}
\exp{\(\frac{[\nabla_1,\nabla_2]dx_1\wedge dx_2}{2\pi i}\)}$ 
where $dx_1\wedge dx_2$ is the formal volume form on the noncommutative 
two-tori $\cA_\theta$, and one gets 
$\rank(E_{g_a,\theta})=|q_a+p_a\theta|$ 
and first Chern class (, \ie , degree) $p_a\frac{q_a+p_a\theta}{|q_a+p_a\theta|}$. 

\begin{defn}
Let $E_a$ be a Heisenberg module $E_{g_a,\theta}$ equipped with 
a constant curvature connection (\ref{ccc}) with 
$(\alpha,\beta)=(\alpha_a,\beta_a)$. 
We define the connection $\nabla_{ab,i}$ on $\Hom(E_a,E_b)$ by 
\begin{equation}
 \nabla_{ab,1}=\ov{q_a+p_a\theta}\fpartial{x}-2\pi i\beta_{ab}\ ,\qquad
 \nabla_{ab,2}=-\ov{q_a+p_a\theta}
\frac{2\pi i p_{ab}x}{q_{ab}+p_{ab}\theta_a}-2\pi i\alpha_{ab}
\end{equation}
for $p_{ab}\ne 0$, and for $p_{ab}=0$, 
\begin{equation}
 \begin{split}
 &\nabla_{aa',1}(U_1)^{n_1}(U_2)^{n_2}
 =2\pi i\(\frac{n_1}{q_a+p_a\theta}-\beta_{aa'}\)(U_1)^{n_1}(U_2)^{n_2}\ ,\\
 &\nabla_{aa',2}(U_1)^{n_1}(U_2)^{n_2}
 =2\pi i\(\frac{n_2}{q_a+p_a\theta}-\alpha_{aa'}\)(U_1)^{n_1}(U_2)^{n_2}\ , 
 \end{split}
\end{equation} 
where 
\begin{equation}
 \alpha_{ab}:=\frac{\alpha_b}{q_b+p_b\theta}
-\frac{\alpha_a}{q_a+p_a\theta}\ ,\qquad 
\beta_{ab}:=\frac{\beta_b}{q_b+p_b\theta}
-\frac{\beta_a}{q_a+p_a\theta}\ .
\end{equation}
 \label{defn:conn-hom}
\end{defn}

 \subsection{The tensor product of Heisenberg modules}
\label{ssec:tensor}

Next, consider a tensor product between 
$\cA_{g_a\theta}$-$\cA_{g_b\theta}$ bimodule and 
$\cA_{g_b\theta}$-$\cA_{g_c\theta}$ bimodule. 
Namely, we construct a map $\varphi$
\begin{equation}
 \varphi: \Hom(E_{g_a,\theta},E_{g_b,\theta})
 \otimes \Hom(E_{g_b,\theta},E_{g_c,\theta})\raw 
 \Hom(E_{g_a,\theta},E_{g_c,\theta})\ . 
 \label{varphi}
\end{equation}
As stated in the previous subsection, 
$\Hom(E_{g_a,\theta},E_{g_b,\theta})$ is equal to 
$\cA_{\theta_a}$ if $p_{ab}=1$ and 
$E_{g_{ab},\theta_a}$ if $p_{ab}\ne 0$. 
The tensor product $\varphi$ in eq.(\ref{varphi}) is defined so that 
$\varphi(\Hom(E_{g_a,\theta},E_{g_b,\theta})a
 \otimes \Hom(E_{g_b,\theta},E_{g_c,\theta}))
=\varphi(\Hom(E_{g_a,\theta},E_{g_b,\theta})
 \otimes a\Hom(E_{g_b,\theta},E_{g_c,\theta}))$ 
for $a\in\cA_{\theta_b}$. 
If $p_{ab}=p_{bc}=0$, this tensor product is just the usual 
product in $\cA_{\theta_a}$. 
If $p_{ab}=0$ and $p_{bc}\ne 0$, it is given by the action of 
$\cA_{\theta_a}=\cA_{\theta_b}$ on $E_{g_{bc},g_b\theta}$ (eq.(\ref{H-mod})). 
In the case $p_{ab}\ne 0$ and $p_{bc}=0$, it is given by 
the right action of $\cA_{\theta_b}=\cA_{\theta_c}$ 
on $E_{g_{ab},g_a\theta}$ (eq.(\ref{Ed})). 
In the case $p_{ab}p_{bc}\ne 0$, 
if $g_{ab}g_{bc}=1$, the tensor product 
$\varphi:E_{g_{ab},\theta_a}\otimes E_{g_{ba},\theta_b}\raw 
\cA_{\theta_a}$ is given by 
\begin{equation}
 \varphi(f_{ab}\otimes
f_{ba})=\sum_{(n_1,n_2)\in\Z^2}
(U_1)^{n_1}(U_2)^{n_2}
\sum_{j\in\Z_{|p_{ab}|}}
 \int dx ((U_2)^{-n_2}(U_1)^{-n_1}f_{ab}(x,j))
 f_{ba}(\frac{x}{q_{ab}+p_{ab}\theta_a},-q_{ab}j)\ .
 \label{tensor-=}
\end{equation}
Using eq.(\ref{dag}), one can also write this as 
\begin{equation*}
 \varphi(f_{ab}\otimes
f_{ba})=\sum_{(n_1,n_2)\in\Z^2}
(U_1)^{n_1}(U_2)^{n_2}
\sum_{j\in\Z_{|p_{ab}|}}
 \int dx ((U_2)^{-n_2}(U_1)^{-n_1}f_{ab}(x,j))
 ((f_{ba}^\dag)^*(x,j))\ .
\end{equation*}
This type of the tensor product is sometimes called an inner product 
in the context of Morita equivalence on noncommutative tori 
(see \cite{Rhigh}). 

Alternatively, for $g_{ab}g_{bc}\ne 1$, 
a tensor product was first constructed explicitly in \cite{Stensor} 
and modified in \cite{foliation} for our purposes.
\footnote{
In \cite{Stensor} the tensor product between right $\cA_\theta$ modules 
and left $\cA_\theta$ modules is constructed. 
In that case, for right $\cA_\theta$ module $E_{g_a,\theta}$ and 
left $\cA_\theta$ modules $E_{g_b,\theta}$, 
the tensor product is defined so that 
$E_{g_a,\theta}U_i\otimes_{\cA_\theta}E_{g_b,\theta}
\sim E_{g_a,\theta}\otimes_{\cA_\theta}U_iE_{g_b,\theta}$ 
where $U_i$'s are two generators of $\cA_\theta$. 
We instead define the tensor product 
as in eq.(\ref{contract}) for our purposes. See also \cite{PoSc}, 
where the tensor product is of the same kind as ours, but in slightly 
different notations. }
It is given by 
\begin{equation}
 \varphi(f_{ab}\otimes f_{bc})(x,j)=
 \sum_{u\in\Z}f_{ab}(x+\ov{p_{ab}}(u-\frac{p_{bc}}{p_{ac}}j),-r_{ab}u+j)
 \cdot
 f_{bc}(\frac{x}{q_{ab}+p_{ab}\theta_a}
 -\frac{q_{bc}+p_{bc}\theta_b}{p_{bc}}(u-\frac{p_{bc}}{p_{ac}}j), u)\ .
 \label{tensor}
\end{equation}
In any case, the tensor product is defined so that it satisfies 
\begin{equation}
 \varphi((f_{ab} Z_i)\otimes f_{bc})=\varphi(f_{ab}\otimes (U_i f_{bc}))
 \label{contract}
\end{equation}
for $i=1, 2$, and further 
\begin{equation}
 \begin{split}
  \varphi((U_i f_{ab})\otimes f_{bc})&=U_i\varphi(f_{ab}\otimes f_{bc})\\
  \varphi(f_{ab}\otimes (f_{bc} Z_i))&=\varphi(f_{ab}\otimes f_{bc})Z_i
 \end{split}
 \label{contract2}
\end{equation}
for $i=1, 2$.  Here $U_i$'s in the left hand side are those which 
act on $E_{g_{ab},\theta_a}$ and $U_i$'s in the right hand side are 
those which act on $E_{g_{ac},\theta_c}$. 
Similarly $Z_i$'s in both sides are different from each other. 
By direct calculations one can check that the tensor product satisfies eq.(\ref{contract}) 
and (\ref{contract2}). 
\begin{lem}
The tensor product $\varphi$ 
is associative. Namely, for any $g_{ab},g_{bc},g_{cd}\in SL(2,\Z)$ and 
$f_{ab}\in\Hom(E_a,E_b)$ etc. , 
\begin{equation*}
 \varphi(\varphi(f_{ab}\otimes f_{bc})\otimes f_{cd})
 =\varphi(f_{ab}\otimes \varphi(f_{bc}\otimes f_{cd}))
\end{equation*}
holds. 
 \label{lem:asso}
\end{lem}
\pf The proof is essentially the same as those given by \cite{PoSc}.  \qed
\begin{lem}
The connection on $\Hom(E_a,E_b)$ in Definition \ref{defn:conn-hom} 
satisfies the Leibniz rule with respect to the tensor product. 
Namely, for any $a,b,c$ 
and $f_{ab}\in\Hom(E_a,E_b)$, $f_{bc}\in\Hom(E_b,E_c)$, 
\begin{equation}
 \nabla_{ac,i}\varphi(f_{ab}\otimes f_{bc})
 =\varphi(\nabla_{ab,i}(f_{ab})\otimes f_{bc})
  +\varphi(f_{ab}\otimes\nabla_{bc,i}(f_{bc})) 
 \label{Leibniz}
\end{equation}
holds. 
 \label{lem:Leibniz}
\end{lem}

Using eq.(\ref{tensor-=}), we introduce two types of inner 
products between Heisenberg modules. One is given by 
the trace of the tensor product (\ref{tensor-=})
\begin{equation}
  \la f_{ab},f_{ba}\ra:=\Tr\varphi(f_{ab}\otimes f_{ba})\ ,
 \label{inner-product}
\end{equation}
where $\Tr$ is the one defined in Definition \ref{defn:trace}. 
Another one is, for $f_{ab},f'_{ab}\in E_{g_{ab},\theta_a}$, 
\begin{equation}
   \la f'_{ab}|f_{ab}\ra:=\Tr\varphi(f_{ab}\otimes (f'_{ab})^\dag)\ .
 \label{inner-product2}
\end{equation}
Using the latter one, one can further define 
a norm. For $f_{ab}\in E_{g_{ab},\theta_a}$ it is defined by 
\begin{equation}
 ||f_{ab}||=(\la f_{ab}|f_{ab}\ra)^\half\ .
 \label{norm}
\end{equation}
\begin{lem}
The inner product $\la\ ,\ \ra$ in (\ref{inner-product}) is 
symmetric, that is, 
\begin{equation}
 \la f_{ba},f_{ab}\ra=\la f_{ab},f_{ba}\ra
 \label{symm-inner}
\end{equation}
holds. 
\end{lem}
\pf
This follows from eq.(\ref{tensor-=}) and the 
definition of the trace (Definition \ref{defn:trace}).

 \subsection{T-duality for open strings 
and mirror symmetry of noncommutative two-tori}
\label{ssec:string}

In this subsection, 
we discuss a physical aspect of our arguments. 
We propose a mirror symmetry for noncommutative two-tori, which is 
the background of our homological mirror symmetry set-up in the next section, 
and explain the relation of noncommutativity with the $t$-structure in D-brane stability.

Heisenberg modules discussed in subsection \ref{ssec:nctori} and \ref{ssec:tensor} 
can be thought of as 
noncommutative analogue of vector bundles over a noncommutative two-torus. 
Physically, a 
Heisenberg module $E_{g,\theta}$ with $\(\bps q & s\\ p & r\eps\)$ 
is identified with a $q$ D2-branes $p$ D0-branes bound state on the noncommutative two-torus. 
Thus, by adding a notion of complex structure, one can consider a noncommutative analogue of 
a category of coherent sheaves on two-tori. 
Though these Heisenberg modules are based on the crossed product expression of 
a noncommutative two-torus $\cA_\theta$, 
a noncommutative two-torus defined by deformation quantization (cf. \cite{Rdiff}) 
may be better understood geometrically, since it reduces to 
a usual commutative two-torus in a commutative limit. 
In this picture, one can construct topologically twisted vector bundles over a noncommutative 
two-torus called quantum twisted bundles (\cite{Ho,MZ,BMZ} and see \cite{KS}). 
Then, a Heisenberg module is in fact isomorphic to a quantum twisted bundle with compatibility 
of the structures of connections and so on (see \cite{MZ,KS}). 
On the other hand, the crossed product representation of the noncommutative two-torus $\cA_\theta$ 
and Heisenberg modules over it have a geometric realization which is directly related to 
the T-dual expression where open strings stretch between D1-branes \cite{foliation}. 
Namely, one can say that the correspondence between the two descriptions are just T-duality. 
We shall give an explanation about it later in this subsection, and here 
we first recall this geometric realization of Heisenberg modules along \cite{foliation}.  

The underlying geometry of 
the crossed product representation of the rotation algebras $\cA_\theta$ is 
the Kronecker foliation\cite{co-book}. 
Let $\Th^2$ be a two-torus with coordinates $(x_1,x_2)$ 
with periodicity $x_1\sim x_1+1$ and $x_2\sim x_2+1$. 
Here we denote it by $\Th^2$ since it is the dual of a two-torus we consider later. 
For the covering space $\wt{\Th^2}\simeq\R^2$, 
by projections $\pi_1:\wt{\Th^2}\raw S^1\otimes\R$ and $\pi_2:\wt{\Th^2}\raw \R\otimes S^1$ 
corresponding to the identifications $x_1\sim x_1+1$ and $x_2\sim x_2+1$, respectively, 
we have $\pi_{12}\wt{\Th^2}=\Th^2$ where $\pi_{12}:=\pi_1\pi_2=\pi_2\pi_1$. 
Let us consider the line 
\begin{equation}
 x_1+\theta x_2=0
 \label{irrslope}
\end{equation}
on the covering space $\wt{\Th^2}$. 
When $\theta$ is irrational, 
the image of the line (\ref{irrslope}) by $\pi_{12}$ fills densely in $\Th^2$. 
The pair of $\Th^2$ and the image of line (\ref{irrslope}) 
is called the Kronecker foliation with irrational slope $\theta$, 
where the image of the line is called the leaf of the Kronecker foliation. 

On the covering space $\wt{\Th^2}$, there are the mirror images of $S^1$ 
which are expressed as $x_2=u_2\in\Z$. 
$U_1=e^{2\pi i x_1}$ is then the generator of functions on this $S^1$. 
The leaf is necessarily transversal to the $S^1$. 
Let us parametrize the leaf as $(x_1, x_2)=(\theta t+x, -t)$, $t\in\R$. 
The point $t=0$ is an intersection point of the leaf and the $S^1$, 
and the leaf then intersects with the $S^1$ in the next time at $t=1$, 
\ie $(x_1,x_2)=(x+\theta,-1)$. 
Correspondingly, we can define a map $o_U:S^1\raw S^1$ by $o_U(x)=x+\theta$, 
and $U_2: C(S^1)\raw C(S^1)$ is regarded as the pullback; 
\begin{equation*}
 (U_2 a)(x)=(o_U^*) a(x)=a(x+\theta)\ .
\end{equation*}
The cycle $S^1$ is identified with a D1-brane. 
An interval along the leaf, whose ends are on the D1-brane, 
is an open string.  We shall justify this identification later. 
The state $(U_1)^{n_1}(U_2)^{n_2}$ is then the open string state 
which moves to $x_1$-direction by momentum $n_1$ and 
winds $n_2$ times around $x_1$-direction.

The situation above can be thought of the one for $E_{g,\theta}$, 
$g=\(\bps q & s \\ p & r\eps\)\in SL(2,\Z)$ with $p=0$. 
Alternatively, a Heisenberg module $E_{g,\theta}$, $p\ne 0$ 
can be realized as follows. 
We first fix a line $L_{base}\ : x_2=0$, whose image by $\pi_1$ is 
the $S^1$ above. We regard it as 
the `base space', and consider 
lines $L_g^j\ : q x_2=p x_1-q j$, $j=0,\cdots,|p|-1$ in $\wt{\Th^2}$ 
'over' the base space. 
More precisely, we regard a function over a pair $(L_{base},L_g^j)$ 
as a 'vector bundle' over the $S^1$. 
For each $j$, the coordinate of $L_g^j$ is identified with that of $L_{base}$ by 
'open strings'. 
Namely, for $x\in\R$ the coordinate of $L_{base}$ such that $(x_1,x_2)=(q j/p+x,0)$, 
we introduce the coordinate on $L_g^j$ by $x\in\R$ such that 
\begin{equation}
(x_1,x_2)=\(\frac{qx}{q+p\theta}+\frac{qj}{p},\frac{px}{q+p\theta}\)\ .
 \label{coordID}
\end{equation}
Here we set the coordinate so that a point $x=0$ is the intersection point of 
$L_{base}$ and $L_g^j$. We call the intersection point the origin of $(L_{base},L_g^j)$. 
By definition, a point $(x_1,x_2)=(q j/p+x,0)$ in $L_{base}$ and a point 
$(x_1,x_2)=(qx/(q+p\theta)+qj/p,px/(q+p\theta)$ in $L_g^j$ 
are the two ends of an interval along the leaf of the foliation. 
Let us consider a function $f^j$ on it so that $f^j(x)=f(x,j)$ and 
$f\in\cS(\R\times\Z_{|p|})$. 
Furthermore, we prepare infinite copies of pairs $(L_{base},L_g^j)$ by 
parallel-transforming them by $\Z^2$, where the origin is at 
$(x_1,x_2)=(q j/p,0)+(u_1,u_2)$, $(u_1,u_2)\in\Z^2$ 
and the coordinate $x$ is defined so that $x=0$ describes the origin. 
To each of them, we associate a function 
$\ti{f}^j(u_1,u_2)$ as $\ti{f}(u_1,u_2;x,j):=\ti{f}^j(u_1,u_2)(x)=f(x,j)$. 
We think of $\ti{f}^j(u_1,u_2)$ a function on a line 
$(x_1,x_2)=(q j/p+u_1+x,u_2)$ when $U_i$'s act on it, whereas 
a function on a line 
$(x_1,x_2)=(q j/p+u_1+qx/(q+p\theta),u_2+px/(q+p\theta))$ when 
$Z_i$'s act. 
For the operation of $U_i$'s, corresponding to the periodicity $x_1\sim x_1+1$ of 
the line $(x_1,x_2)=(q j/p+u_1+x,u_2)$, the period of $x$ is $x\sim x+1$. 
The action of $U_1$ is essentially the multiplication of 
the generator of functions on the line with periodicity $x\sim x+1$, that is, 
$U_1\sim e^{2\pi i x}$. 
However, it is slightly modified so that 
$U_1(u_1,u_2;x+q/p,j)=U_1(u_1',u_2';x,j+1)$ holds 
since $(u_1,u_2;x+q/p,j)$ and $(u_1',u_2';x,j+1)$ denote the same point in $\Th^2$. 
In this way the action of $U_1$ is given by 
\begin{equation*}
  U_1(u_1,u_2;x,j)=e^{2\pi i(x+j\frac{q}{p})}\ 
(\ =U_1(u'_1,u'_2; x-\frac{q}{p},j+1)\ )\ 
\end{equation*}
as multiplication on functions on $\Z^2\times\R\times\Z_{|p|}$. 
Alternatively, 
the action of $U_2$ is defined as the pullback of 
$o_U : \Z^2\times\R\times\Z_{|p|}\raw\Z^2\times\R\times\Z_{|p|}$, which we give by 
$o_U(u_1,u_2; x,j)=(u_1,u_2-1; x+q/p+\theta,j+1)$ so that 
the origin is transformed from $(q j/p+u_1,u_2)$ to 
$(q j/p+u_1,u_2)-(q/p,1)=(q (j-1)/p+u_1,u_2-1)$ and two points 
$(u_1,u_2; x,j)$, $((u_1,u_2-1; x+q/p+\theta,j-1))$ are the two end points of an interval 
(open string) along the leaf. 
Thus, one obtains 
\begin{equation*}
  (U_2\ti{f})(u_1,u_2;x,j)=\ti{f}(u_1,u_2-1;x+\frac{q}{p}+\theta,j-1)\ .
\end{equation*}
On the other hand, for $Z_i$'s, 
the period is $x\sim x +(q+p\theta)$ in the sense that 
$x$ and $x+(q+p\theta)$ describe the same point in $\Th^2$. 
In addition, $(u_1,u_2;x+(q+p\theta)/p,j)$ and $(u'_1,u'_2;x,j+1)$ 
should represent the same point on the D1-brane. Thus, we define 
\begin{equation*}
  Z_1(u_1,u_2;x,j)=e^{2\pi i(\frac{x}{q+p\theta}+\frac{j}{p})}\ .
\end{equation*}
The action of $Z_2$ is then the pullback of 
$o_Z: \Z^2\times\R\times\Z_{|p|}\raw\Z^2\times\R\times\Z_{|p|}$, 
$o_Z(u_1,u_2; x,j)=(u_1-s,u_2; x+1/p,j-r)$ 
which transform the origin $(q j/p+u_1,u_2)$ to $(q j/p+u_1,u_2)-(1/p,0)=(q (j-r)/p+u_1-s,u_2)$ 
and an open string starting at $(u_1,u_2;x,j)$ ends at $(u_1-s,u_2;x+1/p,j-r)$. 
Then, one obtains 
\begin{equation*}
 (\ti{f} Z_2)(u_1,u_2;x,j)=\ti{f}(u_1-s,u_2;x+\ov{p},j-r)\ .
\end{equation*}
Then, projecting $\ti{f}^j(u_1,u_2)$ to $f^j$ by simply dropping indices $(u_1,u_2)$, 
we get the Heisenberg module $E_{g,\theta}$ in subsection \ref{ssec:nctori}. 
One can see that the action of $U_i$ and $Z_i$ for the same $i=1,2$ commutes trivially, 
and the action of $U_i$ and $Z_i$ for different $i$ commutes by 
the definition of the coordinate eq.(\ref{coordID}). 
Note that these realizations of Heisenberg modules admit the ambiguity of 
parallel transformations on $\wt{\Th^2}$, 
which, we see in subsection \ref{ssec:cat}, correspond to $\alpha_a$.

Although we identify a Heisenberg module $E_{g,\theta}$ with a D1-brane winding cycle $(q,p)$, 
it can be identified with the space of open strings 
binding between a $(0,1)$ D1-brane (base $S^1$) and a $(p_a,q_a)$ D1-brane 
if it is treated as a $\cA_\theta$-$\cA_{\theta_a}$ bimodule. 
In more general, 
$\Hom(E_a,E_b)$ is regarded as the space of open strings stretching 
between D1-branes $E_a$ and $E_b$ (see \cite{SW,KMT}). 
Since $\Hom(E_a,E_b)=E_{g_{ab},\theta_a}$ is 
characterized by two lines with slope 
$p_a/q_a$ and $p_b/q_b$ on the covering space, 
this situation can clearly be 
reduced to the case explained above 
by $SL(2,\Z)$ translation $(g_a)^{-1}$ on $\Th^2$. 
The composition of $\Hom$ is then the interaction of open strings 
which is associative.

We called intervals binding between two D1-branes 
along the leaves of the foliation open strings. 
It is in fact justified as follows \cite{foliation}. 
Now the noncommutativity of deformation quantization \cite{BFFLS} 
is well understood from open string theory with 
$B$-field background (see \cite{CF,Scho,SW}). 
Along this line, noncommutative two-tori is obtained by the algebra of 
correlation functions of open strings with both ends on D2-brane 
on a two-torus $\T^2$ \cite{SW}. 
The explanation is simplified if the Seiberg-Witten limit is taken 
\cite{SW}, so we consider this situation. 
Let two by two matrix $E:=g+B$ be the background of the $\T^2$ 
on which a D2-brane exists, 
where $g=\(\bps g_{11} & g_{12}\\ g_{12} & g_{22}\eps\)$ is a positive
definite metric and $B=\(\bps 0 & -b \\ b & 0 \eps\)$ is a shewsymmetric 
matrix called $B$-field.
\footnote{Do not confuse this metric $g$ with $g\in SL(2,\Z)$ 
used to define a Heisenberg module. There are no connection 
between them. } 
It is known that 
in the Seiberg-Witten limit $g\raw 0$ 
a noncommutative torus $\cA_\theta$ with $\theta=-\ov{b}$ is obtained . 

Let us T-dualize for $x_2$-direction. The background $E$ is transformed to $\Eh$ \cite{GPR}, 
\begin{equation*}
 \Eh=(I_1 E+ I_2)(I_2 E + I_1)^{-1}\ ,\qquad 
 I_1:=\bp 1 & 0 \\ 0 & 0\ep,\ \ \ I_2:=\bp 0 &0  \\ 0 & 1\ep\ , 
\end{equation*}
or more explicitly, $\Eh:=\gh+{\hat B}$ is given by 
\begin{equation}
 \gh+{\hat B}
 =g_{22}^{-1}\bp \det{(g)}+b^2 & -b\\ -b & 1\ep 
 +g_{22}^{-1}\bp 0 & g_{12}\\ -g_{12} & 0\ep\ . 
 \label{Tdualbg}
\end{equation}
For open string theory, this T-duality transforms 
a D2-brane on $\T^2$ to a $(1,0) $D1-brane (D1-brane on the base $S^1$). 
In the Seiberg-Witten limit $g\raw 0$, the above metric reduce to 
\begin{equation}
 \gh\sim g_{22}^{-1}b^2
 \bp 1 & -\ov{b}\\
     -\ov{b} & \ov{b^2} \ep\ . 
 \label{deg}
\end{equation}
Open strings ending on the D1-brane take their configuration 
so that they have the minimum masses. 
Here, one can see that this metric 
is degenerate along the leaf with slope $-\ov{\theta}=b$. 
By this T-duality transformation, 
a bound state of $q$ D2-brane and $p$ D0-brane, which we call a $(q,p)$ D2-D0 brane, 
corresponds to a $(q,p)$ D1-brane, a D1-brane on a geodesic cycle 
winding $q$-times around $x_1$-direction and $p$-times around $x_2$-direction. 
Open strings binding between any such two D1-branes 
stretch along the leaf and the situation coincides with the 
noncommutative two-tori associated with the Kronecker foliation. 
Consequently, we can conclude that the two representations of 
the noncommutative two-tori $\cA_\theta$, 
that in the deformation quantization and that 
in the crossed product, are related by the T-duality \cite{foliation}.

For two-tori, this T-duality is equivalent to the mirror symmetry \cite{DVV}. 
The pair of complex structure $\tau$ and complexified symplectic structure
$\rho$ are defined in one-to-one correspondence 
with the backgrounds $E=g+B$ as 
\begin{equation}
 \tau=\frac{g_{12}}{g_{11}}+i\frac{\sqrt{g}}{g_{11}}
\ ,\qquad \rho=i\sqrt{g}+b\ .
 \label{taurho}
\end{equation}
The mirror dual torus is defined so that 
the complex structure and the complexified symplectic structure 
are interchanged. However, we have an ambiguity of the identification which comes from 
an automorphism $SL(2,\Z)$ acting 
compatibly on each two-torus (see \cite{K}). 
One can check that, by our choice of T-duality (\ref{Tdualbg}), the 
complex structure and the complexified symplectic structure 
are transformed as 
\begin{equation}
 {\hat \tau}=-\ov{\rho}\ ,\qquad {\hat \rho}=-\ov{\tau}\ .
 \label{mirror}
\end{equation} 
In this paper we define ${\hat \tau}$ and ${\hat \rho}$ the complex structure 
and the complexified symplectic structure, respectively, 
on the mirror dual torus.
The homological mirror of (commutative) two-tori then claims an equivalence of 
a category $\cC^B$ of coherent sheaves on $\T^2$ (B indicates the corresponding 
topological string theory is the B-model \cite{W2}) and 
the Fukaya's $A_\infty$ category $\cC^A$ on the mirror dual two-torus $\Th^2$ 
(A-model side) for a fixed $\rhoh=-\ov{\tau}$ (and vice visa). 
The category of coherent sheaves depends only on the 
complex structure $\tau$ and is independent of the complexified symplectic 
structure $\rho$, while the $A_\infty$-category depends 
only on the complexified symplectic structure $\rhoh$. 
Especially, for our two-tori case, the coherent sheaves, the objects of $\cC^B$, 
are identified with D2-D0 brane bound states, whereas 
the objects of $\cC^A$ are the D1-branes on the geodesic cycles. 
As stated previously, from a open string physics viewpoint, 
both $\T^2$ and $\Th^2$ are 
associated with noncommutative two-tori $\T_\theta^2$ of the same $\theta$ 
but different descriptions, especially in the Seiberg-Witten limit $g\raw 0$. 
Though the Heisenberg modules are interpreted from the geometry of crossed product representation of 
noncommutative two-torus $\T_\theta^2=\cA_\theta$, 
we treat them as the T-dual side, the B-model on $\T^2_\theta$. 
Namely, by defining a complex and holomorphic structures on $\T_\theta^2$ we can consider 
a noncommutative version of $\cC^B$, which we denote by $\cC^B_\theta$. 
On the other hand, by the T-dual correspondence we can expect what should be taken for the mirror side, 
$\cC^A_\theta$. As a result, it is essentially equivalent to $\cC^A$, the commutative case, 
and some correspondence of $\cC^B_\theta$ and $\cC^A_\theta$ is in fact observed explicitly 
in \cite{foliation}.  
Here, the Seiberg-Witten limit corresponds to the 
boundary of the upper half plane, the moduli of $\rho$. 
The shape of the dual torus is degenerate since $\tauh=\theta$. 
In order to extend the correspondence between closed string variables $(\tau,\rho)$ 
(or $g+B$) 
and the noncommutativity $\theta$ to the general situation $g\ne 0$, 
we need one more parameter $\phi$ so that the action of Morita equivalence acts 
compatibly. The correspondence is then given as follows \cite{PS}; 
\begin{equation}
 (g+B)^{-1}=(G+\Phi)^{-1}+\Theta\ ,
 \label{commNC}
\end{equation}
where $\Phi:=\(\bps 0 & -\phi \\ \phi & 0\eps\)$, $\phi\in\R$ and 
$\Theta:=\(\bps 0 & -\theta \\ \theta & 0\eps\)$. 
If we set $g\raw 0$ and $\phi=0$, we recover the previous situation (\ref{deg}). 
One can also read the identity (\ref{commNC}) as the relation 
between the moduli $g+B$ and $G+\Phi$ for a fixed $\theta$. 
In fact there exists a bijection between the moduli $g+B$ and $G+\Phi$. 
One may identify $G$ and $\Phi$ with the metric and the $B$-field on the 
noncommutative two-torus $\T^2_\theta$, respectively. Thus, 
one can define a complex structure and a complexified symplectic structure 
on $\T^2_\theta$ in a similar way; 
\begin{equation}
 \tau_\theta=\frac{G_{12}}{G_{11}}+i\frac{\sqrt{G}}{G_{11}}
\ ,\qquad \rho_\theta=i\sqrt{G}+\phi\ ,
 \label{taurhoNC}
\end{equation}
and the mirror symmetry for noncommutative two-tori can be defined by ; 
\begin{equation}
 \tauh_\theta=-\ov{\rho_\theta}\ ,\qquad \rhoh_\theta =-\ov{\tau_\theta}\ .
 \label{ncmirror}
\end{equation}
This is in fact a $\Z_2$ symmetry. Thus, the mirror symmetry is extended on the 
one parameter family $\theta$. 
Then, what is the mirror dual noncommutative two-torus ?
It is identified with the (commutative) two-torus with complex structure $\tauh_\theta$ and 
complexified symplectic structure $\rhoh_\theta$ but foliated by lines of 
slope characterized by $\theta$. 
We denote this two-torus by $\Th^2(\tauh_\theta,\rhoh_\theta;\theta)$. 
Let us express them on the complex plane with coordinate $x+iy$, 
where the two-torus is described by periods $(x,y)=(1,0)$ and $(x,y)=(\Re(\tauh_\theta),\Im(\tauh_\theta))$. 
Then the leaf of the foliation is defined by parallel transformations of line $L_\theta$, 
\begin{equation}
 L_\theta : \Im(\tauh_\theta)x+\theta y=0\ .
 \label{Ltheta}
\end{equation}
On the other hand, by a direct calculation one can see that $\tau$ is invariant under 
the change of $\theta$, $\tau_\theta=\tau$. 
Namely, for two-tori, the homological mirror symmetry on this noncommutative set-up should be 
essentially the same as the commutative case. 
The only $\theta$-dependence comes through $t$-structure of the D-brane stability 
\cite{Dou,Bridge}. 
The basic tool in the D-brane stability is the central charge $\cZ$. 
For commutative two-tori, it can be defined by 
\begin{equation}
 \cZ_{\rho}(E_{g_a,\theta=0})=\ov{g_s}\(\rho q_a-p_a\)\ ,
 \label{cc-com}
\end{equation}
where $g_s\in\C$ is the complexified string coupling constant. 
We call $E_{g_a,\theta}$ the D-brane $a$, and sometimes write the corresponding central charge 
as $\cZ_{\rho}(E_{g_a,\theta=0})=\cZ_{\rho}(a)$. 
The absolute value defines the lowest bound of D-brane bound states 
with $\rank=q_a$ and $\deg=p_a$. 
\begin{equation}
 |\cZ_{\rho}(a)|=\frac{|q_a|}{|g_s|}\sqrt{\det\(g+B-\frac{p_a}{q_a}\)}\ .
\end{equation}
This is nothing but the mass of the D-brane with constant curvature connections. 
Alternatively, in the variables of the dual torus $\Th^2(\tauh_\theta,\rhoh_\theta;\theta)$, 
one can see that the central charge (\ref{cc-com}) is expressed as 
\begin{equation}
\cZ_{\tauh}(a) \propto \(q_a+\tauh p_a\)\ . 
\end{equation}
In an appropriate definition of the complexified string coupling, 
the absolute value again gives the mass of $(q,p)$ D1-brane 
on $\Th^2(\tauh_\theta,\rhoh_\theta;\theta)$. 
The angle of $q_a+\tauh p_a$ is just the angle of D1-brane $a$. 
If we consider the sum $\cZ_{\tauh}(a)+\cZ_{\tauh}(b)$ corresponding to the direct sum of 
the Heisenberg modules $a$ and $b$, in the dual D1-brane picture it is described by 
two segments passing through $0$, $q_a+\tauh p_a$ and $(q_a+q_b)+\tauh(p_a+p_b)$. 
Of course the D1-brane of $\rank=q_a+q_b$, $\deg=p_a+p_b$ with minimum mass is described by 
the straight segment between $0$ and $(q_a+q_b)+\tauh(p_a+p_b)$.
Thus, for 
$n_c(q_c,p_c)=(q_a+q_b,p_a+p_b)$ with $n_c\in\Z$ and relatively prime integers $(q_c,p_c)$, 
we get 
\begin{equation}
|\cZ_{\rho}(a)|+ |\cZ_{\rho}(b)| \ge |\cZ_{\rho}(a)+\cZ_{\rho}(b)| =n_c|\cZ_{\rho}(c)|\ .
 \label{Sbound}
\end{equation}
This is just the stability bound. 

We can extend these central charges on one parameter $\theta$ by the relation (\ref{commNC}) 
with preserving all theses structures. 
The central charge is defined on $\T^2_\theta$ by 
\begin{equation}
 \cZ_{\rho_\theta}(E_{g_a,\theta})=\ov{G_s}\(\rho_\theta q_a-p_a\)\ ,
 \label{cc}
\end{equation}
where $G_s\in\C$ is the complexified (effective) string coupling constant. 
In this definition, the absolute value takes the following form, 
\begin{equation}
 |\cZ_{\rho_\theta}(E_{g_a,\theta})|=\Tr_{\cA_\theta}\ov{|G_s|}\sqrt{\det\(G+\Phi-\frac{p_a}{q_a}\)}
\end{equation}
which is in fact independent of $\theta$ and so especially equal to $\cZ_\rho(E_{g_a,\theta=0})$. 
Namely, one can define the effective string coupling constant $G_s$ so that 
the D-brane masses are preserved for all $p$ and $q$. 
This is just the D-brane mass discussed in the context of noncommutative solitons in \cite{KMT} 
(for solitons on noncommutative two-tori see also \cite{KraSch}). 
Then, the stability bound condition (\ref{Sbound}) remains to hold. 
The equality holds only if $p_a/q_a=p_b/q_b$ for general $\rho$. In this case, 
$E_c$ is called {\it semi-stable}.  
However notice that, in the degenerate limit $g\raw 0$, \ie , $\Im(\rho)\raw 0$, 
we have $|\cZ_{\rho}(a)|+ |\cZ_{\rho}(b)| = |\cZ_{\rho}(a)+\cZ_{\rho}(b)| $ for all $a$ and $b$. 
This means, any D-brane can be decomposed into smaller D-branes. 
This phenomena is just the instability argued in \cite{BKMT}. 
Note that $E_{g,\theta}$ and $E_{-g,\theta}$, in the dual side $\Th^2$, 
describe the D1-branes with the same angles. 
However, they have the D-brane charges of opposite signs, $(q,p)$ and $(-q,-p)$. 
They are relatively distinguished, one is a D-brane, and another one is an anti-D-brane. 
Namely, for a fixed $E_{g,\theta}$, an open string starting from it 
distinguishes whether the endpoint belong to $E_{g',\theta}$ or $E_{-g',\theta}$. 
Let us define the slope of D-brane $a$ by 
\begin{equation}
 -\pi <\Arg(a)\le \pi\ ,\qquad
 \Arg(a):=\Im\log\(\frac{\cZ_{\tauh_\theta}(a)}{\cZ_{\tauh_\theta}(E_{g=1,\theta})}\)
 =\Im\log (q_a+\tauh_\theta p_a) \ .
\end{equation}
In particular $\Arg(a)=0$ for $g_a=1$. 
Since whether $a$ is a D-brane or an anti-D-brane is distinguished only relatively now, 
we need to label the distinction by hand. This is the $t$-structure 
(for more general set-up see \cite{Bridge}). 
It should be good to identify the $t$-structure with noncommutativity $\theta$ ; 
let us define the set of D-branes (not anti-D-branes ) by 
\begin{equation}
 \Ob(\cC_\theta):=\{ a\ |  \Arg(-L_\theta)<\Arg(a)\le\Arg(L_\theta) \}\ ,
\end{equation}
where, for the line $L_\theta$ in eq.(\ref{Ltheta}) we set $\Arg(-L_\theta)$ and $\Arg(L_\theta)$ by 
\begin{equation*}
-\pi<\Arg(-L_\theta):=\Im\log (-i\Im(\tauh_\theta)+\theta)<0\ ,\qquad 
0<\Arg(L_\theta):=\Im\log (i\Im(\tauh_\theta)-\theta)<\pi\ . 
\end{equation*}
In particular, for $\Th_\theta^2$ of $\phi=\theta=0$, we have 
$\Arg(-L_\theta)=-\pi/2$ and $\Arg(L_\theta)=\pi/2$. 
%
%\begin{equation*}
% H^i(\Hom(E_a[k],E_b[l]))=H^{i+l'-k'}(\Hom(E_a[k-k'],E_b[l-l']))\ ,
%\end{equation*}
The important point is that, for a fixed background $g+B$, the set 
$\Ob(\cC_\theta^A)\ (=\Ob(\cC_\theta^B))$ is independent of $\theta$. 
Namely, $\Th^2(\tauh_\theta,\rhoh_\theta;\theta)$ with different $\theta$ can be 
related by a $GL(2,\R)_+$ transformation. 
This is the consequence obtained by translating the argument in \cite{I} into 
our set-up. 
In this way, a noncommutative deformation is understood as a one-parameter extension of the mirror 
symmetry set-up preserving the structure of the corresponding 'D-brane categories' 
if we fix the background $g+B$, whereas it can be regarded as a deformation of $t$-structures 
if we fix the noncommutative background $G+\Phi$. 
In the next section, we concentrate on the case $\phi=0$ for simplicity, 
and propose the corresponding homological mirror set-up as a correspondence between 
a $\theta$-deformed Fukaya category and a category of Heisenberg modules equipped with 
holomorphic structures. 
Since $\tau_\theta$ is independent of $\theta$ as stated previously, 
hereafter we drop it and write simply as $\tau_\theta=\tau$, hence $\rhoh_\theta=\rhoh$.

 \section{On homological mirror on noncommutative two-tori}
\label{sec:ncmirror}

In this section we introduce a complex structure 
on noncommutative two-tori and discuss a noncommutative extension of 
homological mirror symmetry 
on (commutative) two-tori \cite{mirror,PZ,Poli1,Poli2} 
which has its background in the physics in the previous subsection.

In subsection \ref{ssec:hol} we recall the holomorphic structure on 
Heisenberg modules introduced by A.~Schwarz \cite{Stheta,Stensor} 
and construct a cyclic differential graded (DG) category of 
(stable) holomorphic vector bundles over a noncommutative two-torus 
partially based on the work \cite{PoSc}. 
In subsection \ref{ssec:cat} we propose a candidate of the category 
on A-model side, 
the Fukaya's $A_\infty$-category for noncommutative two-tori, 
and set up a homological mirror symmetry conjecture on 
noncommutative two-tori. 
As a part of the conjecture, we prove a categorical mirror symmetry. 
Finally in subsection \ref{ssec:main} 
we complete the Feynman rule to generate higher Massey products, 
which form a cyclic $A_\infty$-category, from the cyclic 
DG category of holomorphic vector bundles on noncommutative two-tori.

 \subsection{Holomorphic structure on noncommutative two-tori} 
\label{ssec:hol}

We have seen that the Heisenberg modules are equipped the constant 
curvature connection as in eq.(\ref{ccc}). 
Now, on noncommutative two-tori 
let us introduce a complex structure $\tau\in \C$ and 
and consider the following combination 
\begin{equation}
\nabb_a:=
\nabla_{a,1}-\ov{\tau}\nabla_{a,2}\ 
\end{equation}
according to \cite{Stheta,Stensor}. 
$\nabb_a$ is regarded as the noncommutative analogue of 
the holomorphic structure. 
The solutions of the following equation
\begin{equation*}
 \nabb_a f(x,j)=0\ 
\end{equation*}
exists iff $\frac{p_a}{q_a+p_a\theta}>0$, and they are given by 
\begin{equation}
 f(x,j)=a^j\exp{\(\frac{2\pi i p_a}{q_a+p_a\theta}
\Big[\frac{\rhoh}{2}\(x+\frac{\alpha_a}{p_a}\)^2
+\frac{\beta_a}{p_a}\(x+\frac{\alpha_a}{p_a}\)\Big]\)}\ , \qquad a^j\in\C\ ,
 \label{holvect2}
\end{equation}
where we denoted $-\ov{\tau}=:\rhoh$. 
They are called theta vectors or holomorphic vectors, since they span a basis of 
a $|p|$-dimensional vector space in $E_{g_a,\theta}$ \cite{Stensor}. 
One can see that $E_{g_a,\theta}$ and $E_{-g_a,\theta}$ is equivalent, in the sense 
that the holomorphic vector (\ref{holvect2}) is invariant under the replacement of 
$(q_a,p_a,\alpha_a)$ with $(-q_a,-p_a,-\alpha_a)$. 
Hereafter, in a reason discussed in the previous subsection 
we take Heisenberg modules $E_{g_a,\theta}$ with $q_a+p_a\theta>0$. 

We can define a holomorphic structure $\nabb_{ab}:\Hom(E_a,E_b)\raw \Hom(E_a,E_b)$ 
in a similar way;  
\begin{equation}
\nabb_{ab}: =
\nabla_{ab,1}+\rhoh\nabla_{ab,2} \ .
 \label{nabb-hom}
\end{equation}
Based on this structure, we construct a cyclic differential 
graded (DG) category of holomorphic vector bundles below. 
A cyclic DG category is a natural extension of a cyclic differential 
graded algebra (DGA) and defined as follows. 
\begin{defn}[Cyclic differential graded category]
A differential graded (DG) category 
$\cC$ consists of a set of objects $\Ob(\cC)=\{a, b, \cdots\}$, 
a $\Z$-graded vector space $V_{ab}=\oplus_k V_{ab}^k$ for each two objects 
$a$, $b$ and grading $k\in\Z$, a degree one differential 
$d :V^k_{ab}\raw V^{k+1}_{ab}$ and a composition of 
morphisms $\varphi: V^k_{ab}\otimes V^l_{bc}\raw V^{k+l}_{ac}$ 
satisfying the following relations ; 
\begin{align}
 & (d)^2=0\ ,\label{diff}\\
 &  d\varphi(v_{ab}\otimes v_{bc})
 =\varphi(d(v_{ab})\otimes v_{bc})
 +(-1)^{|v_{ab}|}\varphi(v_{ab}\otimes d(v_{bc}))\ , \label{diff-leib}\\
 & \varphi(\varphi(v_{ab}\otimes v_{bc})\otimes v_{cd})
=\varphi(v_{ab}\otimes\varphi(v_{bc}\otimes v_{cd}))\ , \label{asso-cat}
\end{align}
where $|v_{ab}|$ is the degree of $v_{ab}$, that is, $v_{ab}\in V_{ab}^{|v_{ab}|}$. 
Let $\eta$ be a nondegenerate symmetric inner product 
of fixed degree $|\eta|\in\Z$ on $V:=\oplus_{a,b}V_{ab}$. 
Namely, for each $a$ and $b$, 
\begin{equation}
 \eta: V^k_{ab}\otimes V^l_{ba}\raw \C
\end{equation}
is nondegenerate, nonzero only if $k+l+|\eta|=0$, and satisfies 
$\eta(V^k_{ab},V^l_{ba})=(-1)^{kl}\eta(V^l_{ba},V^k_{ab})$. 
In this situation, we call a DG category with inner product $\eta$ 
a {\it cyclic DG category} $\cC$ if the following conditions hold; 
\begin{align}
 &\eta(dv_{ab},v_{ba})+(-1)^{|v_{ab}|}\eta(v_{ab},dv_{ba})=0\ ,
 \label{d-int}\\
 &\eta(\varphi(v_{ab}\otimes v_{bc}),v_{ca})
 =(-1)^{{|v_{ab}|}(|v_{bc}|+|v_{ca}|)}\eta(\varphi(v_{bc}\otimes v_{ca}),v_{ab})\ .
 \label{cyclic}
\end{align}
Also, we call a cyclic DG category 
$\cC$ with $d=0$ a {\it cyclic graded category}. 
 \label{defn:cDGcat}
\end{defn}
\begin{rem}
A (cyclic) DG category $\cC$ consisting of one object only is called a (cyclic) 
differential graded algebra (DGA). 
On the other hand, if the space of morphisms $V=\oplus_{a,b}V_{ab}$ is thought of as a 
$\Z$-graded vector space, $(V, d,\varphi)$ and $(V,d,\varphi,\eta)$ can be regarded as 
a DGA and a cyclic DGA, respectively. 
Hereafter we sometimes express explicitly as $(V, d,\varphi)$ or $(V,d,\varphi,\eta)$ 
and call it a (cyclic) DG category. Similarly, we denote a cyclic graded category 
by triple $(V,\varphi,\eta)$. 
 \label{rem:DGcat-alg}
\end{rem}
\begin{prop}[Cyclic DG category of holomorphic vector bundles]
Consider a set $\{a,b,\cdots\}$ where each element $a=(g_a,\alpha_a,\beta_a)$ 
is a Heisenberg module with a constant curvature connection $E_a$ 
(see Definition \ref{defn:conn-hom}) satisfying $q_a+p_a\theta>0$. 
For each pair $(a,b)$, we define a graded vector space of degree zero by 
$V_{ab}^0=\Hom(E_{g_a,\theta},E_{g_b,\theta})$, a graded vector space of degree one by 
$V_{ab}^1=\Hom(E_{g_a,\theta},E_{g_b,\theta})\otimes d\zb$, 
where $d\zb$ is a formal degree one base of the antiholomorphic 1-forms on 
the noncommutative two-tori $\cA_\theta$, 
and $V_{ab}^k=0$ for $k\ne 0,1$. 
On the graded vector space $V=\oplus_{a,b}\oplus_k V_{ab}^k$, 
a differential is defined by 
\begin{equation}
dv_{ab}:=(\nabb_{ab}v_{ab})\otimes d\zb\ . 
\end{equation}
Also, extending the tensor product $\varphi$ 
to $\otimes d\zb$ trivially, one gets a tensor product  
$\varphi:V_{ab}\otimes V_{bc}\raw V_{ac}$. 
An inner product $\eta: V^k_{ab}\otimes V^l_{ba}\raw \C$ is given by 
\begin{equation}
 \eta=\Tr\int dz\varphi\  
 \label{inner-cat}
\end{equation} 
where we set $\int dzd\zb=1$, so 
$\eta(V_{ab}^k,V_{ba}^l)$ is nonzero only if $k+l=1$. 
Then $(V,d,\varphi,\eta)$ forms a 
cyclic DG category. 
 \label{prop:cDGcat}
\end{prop}
\pf 
Since $(d\zb)^2=0$, it is clear that $d$ is a differential
(\ref{diff}). 
The Leibniz rule (\ref{diff-leib}) and the associativity 
(\ref{asso-cat}) just follow from the 
Leibniz rule for $\nabla_{ab,i}$ in Lemma \ref{lem:Leibniz} 
and associativity in Lemma \ref{lem:asso}, respectively, 
in subsection \ref{ssec:tensor}. 
By the definition of the inner product (\ref{inner-cat}), 
it is in fact nondegenerate and symmetric due to
eq.(\ref{symm-inner}). 
The condition (\ref{d-int}) then follows from 
Leibniz rule (Lemma \ref{lem:Leibniz}), and the cyclicity 
(\ref{cyclic}) follow from Lemma \ref{lem:Leibniz} and eq.(\ref{symm-inner}). 
\qed 

\begin{defn}
For each objects $a$ we define a number 
\begin{equation}
 \mu_a=\frac{p_a}{q_a+p_a\theta}\in\R \ . 
\end{equation}
Furthermore, for each $a$ and $b$ we set 
\begin{equation}
 \mu_{ab}:=\frac{p_{ab}}{q_{ab}+p_{ab}\theta_a}
\end{equation}
which characterizes properties of $\Hom(a,b)$. 
 \label{defn:mu}
\end{defn}
\begin{rem}
We have the following identity 
\begin{equation}
 \mu_b-\mu_a=(q_a+p_a\theta)^{-2}\mu_{ab}\ ,
 \label{key}
\end{equation}
so the sign of $\mu_b-\mu_a$ coincides with the sign of $\mu_{ab}$. 
 \label{rem:mu}
\end{rem}
The dimension of the cohomologies $H^0$ and $H^1$ 
are given as follows \cite{PoSc}. 
\begin{itemize}
 \item For $\mu_{ab}>0$, 
 $\dim H^0(V_{ab})=|p_{ab}|$ and $\dim H^1(V_{ab})=0$. 
 \item For $\mu_{ab}<0$, 
 $\dim H^0(V_{ab})=0$ and $\dim H^1(V_{ab})=|p_{ab}|$. 
 \item For $\mu_{ab}=0$, 
 $\dim H^0(V_{ab})=\dim H^1(V_{ab})=0$ for $a\ne b$ and 
 $\dim H^0(V_{ab})=\dim H^1(V_{ab})=1$ for $a=b$. 
\end{itemize}

The cohomology $H^0(V_{ab})$ and $H^1(V_{ab})$ are the kernel and the 
cokernel of $d$, respectively. 
For $\mu_{ab}>0$, 
since $H^0(V_{ab})$ is the kernel of 
\begin{equation*}
 \nabb_{ab}=
\ov{q_a+p_a\theta}
\(\fpartial{x}-\frac{2\pi i\rhoh p_{ab}x}{q_{ab}+p_{ab}\theta_a}\) 
-2\pi i(\beta_{ab}+\rhoh\alpha_{ab})\ ,
\end{equation*}
its base $|0\ra_{ab}^j$, $j=0,\cdots,|p_{ab}|-1$ is given by 
\begin{equation}
 |0\ra_{ab}^j(x):=
 \exp{\(\frac{2\pi i p_{ab}}{q_{ab}+p_{ab}\theta_a}
 \Big[\frac{\rhoh}{2}\(x+\frac{q_b+p_b\theta}{p_{ab}}\alpha_{ab}\)^2
 +\frac{q_b+p_b\theta}{p_{ab}}
 \beta_{ab}\(x+\frac{q_b+p_b\theta}{p_{ab}}\alpha_{ab}\)\Big]\)}\ ,
 \label{vac}
\end{equation}
whose norm (defined by eq.(\ref{norm})) is 
$(\frac{q_{ab}+p_{ab}\theta_a}{T p_{ab}})^{\ov{4}}$, 
where $T=-i(\rhoh-\rhoh^*)=2\Im(\rhoh)$.  

In case $\mu_{ab}<0$, 
$H^0=0$ but $H^1$ is a $|p|$ dimensional vector space. 
Using the inner product, we can now take the cohomology element 
explicitly. 

For $\nabb_{ab}$ in eq.(\ref{nabb-hom}), define 
\begin{equation}
\nabb_{ab}^\dag
= \ov{q_a+p_a\theta}
\(-\fpartial{x}+\frac{2\pi i\rhoh^* p_{ab}x}{q_{ab}+p_{ab}\theta_a}\) 
 +2\pi i(\beta_{ab}+\rhoh^*\alpha_{ab})\ . 
 \label{anti-nabb-hom}
\end{equation}
By the definition of the inner product, we get ;
\begin{lem}
For $f,f'\in E_{g_{ab},\theta_a}$, 
\begin{equation}
 \la f'| \nabb f \ra= \la \nabb^\dag f'| f\ra \ ,\qquad 
 \la f'| \nabb^\dag f \ra= \la \nabb f'| f\ra
 \label{inner2-dag}
\end{equation}
hold. 
 \label{lem:inner2-dag}
\end{lem}
Thus, the cokernel of $V_{ab}^1$ is defined by the states 
annihilated by $\nabb^\dag$. 
It is spanned by $|0\ra_{ab}^j\otimes d\zb$ with 
\begin{equation}
 |0\ra_{ab}^j(x):=
 \exp{\(\frac{2\pi i p_{ab}}{q_{ab}+p_{ab}\theta_a}
 \Big[\frac{\rhoh^*}{2}\(x+\frac{q_b+p_b\theta}{p_{ab}}\alpha_{ab}\)^2
 +\frac{q_b+p_b\theta}{p_{ab}}
 \beta_{ab}\(x+\frac{q_b+p_b\theta}{p_{ab}}\alpha_{ab}\)\Big]\)}\ . 
 \label{anti-vac}
\end{equation}

Now we have 
\begin{equation}
 [\nabb_{ab},\nabb_{ab}^\dag]=
 \frac{2\pi Tp_{ab}}{(q_a+p_a\theta)(q_b+p_b\theta)}\ .
 \label{comm}
\end{equation}
By employing this commutation relation, 
the space $V_{ab}$ is expanded by the Hermite polynomials as follows (cf.\cite{MoPo,PoSc}). 
For $V_{ab}^k$ with $\mu_{ab}>0$, 
define the $|p_{ab}|$-degenerate vacuum states $|0,k\ra_{ab}^j$ 
by those annihilated by $\nabb_{ab}$. We set 
\begin{equation}
 |0,0\ra_{ab}^j=|0\ra_{ab}^j\ ,\qquad
  |0,1\ra_{ab}^j=|0\ra_{ab}^j\otimes d\zb\ .
\end{equation}
Then, $V_{ab}^k$ is spanned by 
\begin{equation*}
 |n,k\ra_{ab}^j:=(\nabb^\dag)^n |0,k\ra_{ab}^j\ ,
\end{equation*}
and $|n,k\ra_{ab}^j$ is expressed by the Hermite polynomials ; 
\begin{equation}
 \begin{split}
& |n,k\ra_{ab}^j(x)= 
 \(\frac{\pi T p_{ab}}{(q_a+p_a\theta)(q_b+p_b\theta)}\)^{\frac{n}{2}}
 H_n\(\sqrt{\frac{\pi T p_{ab}}{q_{ab}+p_{ab}\theta_a}}
\(x+\frac{q_b+p_b\theta}{p_{ab}}\alpha_{ab}\)\)
|0,k\ra_{ab}^j(x)\ , \\
& H_n(x):=\(-\fpartial{x}+2x\)^n\cdot 1\ .
 \end{split}\label{hermite}
\end{equation}

Alternatively, $V_{ab}^k$ with $\mu_{ab}<0$ 
are spanned by the states 
$|n,k\ra_{ab}^j:=(\nabb)^n|0,k\ra_{ab}^j$ where the vacuum states 
are defined by $|0,k\ra_{ab}^j=|0\ra_{ab}^j\otimes d\zb^{\otimes k}$ 
with $|0\ra_{ab}^j$ in eq.(\ref{anti-vac}). 
Just as the Hermite polynomial expansion in eq.(\ref{hermite}), 
each state is written as 
\begin{equation}
 |n,k\ra_{ab}^j(x)= 
 \(\frac{-\pi T p_{ab}}{(q_a+p_a\theta)(q_b+p_b\theta)}\)^{\frac{n}{2}}
 H_n\(\sqrt{\frac{-\pi T p_{ab}}{q_{ab}+p_{ab}\theta_a}}
\(x+\frac{q_b+p_b\theta}{p_{ab}}\alpha_{ab}\)\)
|0,k\ra_{ab}^j(x)\ .
 \label{anti-hermite}
\end{equation}
Thus, each $V_{ab}^k$ is spanned by the linear combination of 
these states, and especially the cohomology states are spanned by 
eq.(\ref{vac}) and (\ref{anti-vac}).

\begin{rem}
By comparing 
$|0\ra_{ab}^j$ in eq.(\ref{vac})and (\ref{anti-vac}), 
and also eq.(\ref{hermite}) and (\ref{anti-hermite}), 
one can see that $|n,k\ra_{ab}^j$ with $\mu_{ab}>0$ and 
$|n,k\ra_{ab}^j$ with $\mu_{ab}<0$ is related by 
the interchange of $\rhoh$ and $\rhoh^*$. 
This interchange leads the interchange of $T$ and $-T$. 
 \label{rem:tau-tau^*}
\end{rem}

Finally, the case of $a,a'$ with $p_{aa'}=0$ is easy. 
In this case we have 
$V_{aa'}^k\simeq \cA_{\theta_a}\otimes (d\zb)^{\otimes k}$, which is spanned by 
\begin{equation}
 |n,0\ra_{aa'} =(U_1)^{n_1}(U_2)^{n_2}\ ,\qquad 
 |n,1\ra_{aa'} =(U_2)^{-n_2}(U_1)^{-n_1}\otimes d\zb\ , 
\end{equation}
where $n=(n_1,n_2)\in\Z^2$. By definition we get 
$(|n,0\ra_{aa'})^\dag\otimes d\zb=|n,1\ra_{aa'}$. 
Since we have 
\begin{equation*}
 \nabb_{aa'}(U_1)^{n_1}(U_2)^{n_2}
 =2\pi i\(\(\frac{n_1}{q_a+p_a\theta}-\beta_{aa'}\)
 +\rhoh\(\frac{n_2}{q_a+p_a\theta}-\alpha_{aa'}\)\)
 (U_1)^{n_1}(U_2)^{n_2}\ ,
\end{equation*}
it is clear that only when $a=a'$ the cohomology is nonzero and 
$\dim H^k(V_{aa})=1$ for $k=0,1$. 
We denote these basis by 
$1=:|0,0\ra_{aa}$ and $1\otimes d\zb=:|0,1\ra_{aa}$. 
\begin{rem}
For any $a$ and $b$, the compatibility of the inner product with $\nabb$, $\nabb^\dag$ 
leads the following relation  
\begin{equation}
  \(|n,k\ra_{ab}^j\)^\dag= |n,k\ra_{ba}^{-q_{ab}j}\  .
\end{equation}
\end{rem}
{}Comparing this with eq.(\ref{inner-product2}), we have the following. 
\begin{cor}
Any element $f_{ab}\in V_{ab}^k$ is expanded as 
\begin{equation}
 f_{ab}=\sum_{n}|n,k\ra_{ab}^j N_{ab}(n)
 \eta(f_{ab},|n,1-k\ra_{ba}^{-q_{ab}j})\ ,
\end{equation}
where $N_{ab}(n)$ is the normalization and given by 
\begin{equation}
 \begin{split}
  \sqrt{\frac{Tp_{ab}}{(q_a+p_a\theta)(q_b+p_b\theta)}}
\ov{[\nabb_{ab},\nabb_{ab}^\dag]^n n!} &\quad\mbox{for}\ \ \mu_{ab}>0\ , \\
  \sqrt{\frac{Tp_{ab}}{(q_a+p_a\theta)(q_b+p_b\theta)}}
\ov{[\nabb_{ab},\nabb_{ab}^\dag]^n n!} &\quad\mbox{for}\ \ \mu_{ba}<0\ , \\
   \ov{q_a+p_a\theta}
 &\quad\mbox{for}\ \ \mu_{aa'}=0\ .
 \end{split}
 \label{norma}
\end{equation}
\end{cor}
\begin{defn}[Hodge-Kodaira decomposition]
Now, for any $a,b$, $V_{ab}^k$ is spanned by $|n,k\ra$, where 
$n\in\Z_{\ge 0}$ for $\mu_{ab}\ne 0$ and $n=(n_1,n_2)\in\Z^2$ for $\mu_{ab}=0$. 
In any case, we have $|n,k\ra=0$ if $k\ne 0,1$. 
We define $d^{-1}$ by 
\begin{equation*}
 d^{-1}|n,k\ra_{ab}^j=\ov{(n+1)[\nabb_{ab},\nabb_{ab}^\dag]}
 |n+1,k-1\ra_{ab}^j
\end{equation*}
for $\mu_{ab}>0$, 
\begin{equation*}
 d^{-1}|n,k\ra_{ba}^j=|n-1,k-1\ra_{ba}^j\  (n\ge 1)\ ,
 \qquad d^{-1}|0,k\ra_{ba}^j=0 
\end{equation*}
for $\mu_{ba}<0$, and 
\begin{equation*}
 d^{-1}|n,k\ra_{aa'}
 =\ov{2\pi i\(\(\frac{n_1}{q_a+p_a\theta}-\beta_{aa'}\)
+\rhoh\(\frac{n_2}{q_a+p_a\theta}-\alpha_{aa'}\)\)}|n,k-1\ra_{aa'}
\end{equation*}
for $\mu_{aa'}=0$. 
Here if $a=a'$, we set $d^{-1}|0,1\ra=0$. 
Then we get the Hodge-Kodaira decomposition 
\begin{equation}
dd^{-1}+d^{-1}d+P=\1
\end{equation}
for each graded vector space $V_{ab}$. 
In particular, on $V_{ab}^0$ one gets $d^{-1}d+P=\1$ since $d^{-1}=0$, 
and on $V_{ab}^1$ one gets $dd^{-1}+P=\1$ since $d=0$. 
 \label{defn:HK}
\end{defn}

 \subsection{Categorical mirror symmetry on 
noncommutative two-tori}
 \label{ssec:cat}

We set up the homological mirror symmetry conjecture on noncommutative
two-tori as an equivalence between two (cyclic) $A_\infty$-categories. 
Just as a DG category is a natural extension of a DGA, 
an $A_\infty$-category is a natural extension of an 
$A_\infty$-algebra introduced by J.~Stasheff \cite{Sta1,Sta11}. 
\begin{defn}[(Cyclic) $A_\infty$-category \cite{Fukaya}]
An $A_\infty$-category $\cC$ consists of a set of objects $\Ob(\cC)=\{a,b,\cdots\}$, 
$\Z$-graded vector space $\Hom(a,b)$ 
for each pair of objects $a,b\in\Ob(\cC)$, 
and a collection of degree one multilinear compositions 
$\m:=\{m_n:\Hom(a_1,a_2)\otimes\Hom(a_2,a_3)\otimes
\cdots\otimes\Hom(a_n,a_{n+1})\raw\Hom(a_1,a_{n+1})\}_{a_i\in\Ob(\cC),\ n\ge 1}$
satisfying the following relations
\begin{equation}
 \begin{split}
&0=\sum_{k+l=n+1}\sum_{j=0}^{k-1}
(-1)^{|e_{12}|+\cdots+|e_{j(j+1)}|}\\
&\ m_k(e_{12},\cdots,e_{j(j+1)},m_l(e_{(j+1)(j+2)},\cdots,e_{(j+l)(j+l+1)}),
 e_{(j+l+1)(j+l+2)},\cdots,e_{n(n+1)}) \
 \end{split}
 \label{Ainftyrel}
\end{equation}
for $n\ge 1$. 
Here $|e_{i(i+1)}|$ on $(-1)$ denotes the degree of 
$e_{i(i+1)}\in\Hom(a_i,a_{i+1})$. 
One can also express these relations as 
$0=\sum_{k+l=n+1}\sum_{j=0}^{k-1}
m_k(\1^{\otimes j}\otimes m_l\otimes\1^{\otimes (n-j-l)})
(e_{12},\cdots,e_{n(n+1)})$. 
Moreover, we call $\cC$ a {\it cyclic $A_\infty$-category} 
if it has a nondegenerate skew-symmetric inner product of fixed integer degree, 
\begin{equation*}
 \omega: \Hom(a,b)\otimes\Hom(b,a)\raw\C\ ,
\end{equation*}
satisfying the following cyclic condition 
\begin{equation*}
 \omega(m_n(e_{12},\cdots,e_{n(n+1)}),e_{(n+1)1})
 =(-1)^{(|e_{23}|+\cdots +|e_{(n+1)1}|)|e_{12}|}
 \omega(m_n(e_{23},\cdots,e_{(n+1)1}),e_{12})\ . 
\end{equation*}
For $\deg(\omega)=|\omega|$, the sign in the equation above can also be written as 
$(-1)^{(|e_{23}|+\cdots +|e_{(n+1)1}|)|e_{12}|}=(-1)^{(-|\omega|-1-|e_{12}|)|e_{12}|}
=(-1)^{|\omega| |e_{12}|}$. 
 \label{defn:Ainfty}
\end{defn}
\begin{rem}
An $A_\infty$-category $\cC$ consisting of one object only is equivalent to an 
$A_\infty$-algebra \cite{Sta1,Sta11}. 
On the other hand, regarding $\cH=\oplus_{a,b}\cH_{ab}$, 
$\cH_{ab}:=\Hom(a,b)$ as a $\Z$-graded vector space, 
we can think of $(\cH,\m)$ as an $A_\infty$-algebra, and $(\cH,\omega,\m)$ 
as a cyclic $A_\infty$-algebra (for cyclic $A_\infty$-algebras see \cite{thesis}). 
Hereafter we sometimes denote a (cyclic) $A_\infty$-category 
explicitly by $(\cH,\m)$ or $(\cH,\omega,\m)$. 
 \label{rem:cat-alg}
\end{rem}
\begin{defn}[Minimal $A_\infty$-category]
An (cyclic) $A_\infty$-category $\cC$ 
is called {\it minimal} if $m_1=0$. 
 \label{defn:minimal}
\end{defn}
This corresponds to that $(\cH,\m)$ with $m_1=0$ is a minimal $A_\infty$-algebra. 
\begin{defn}[(Cyclic) $A_\infty$-functor]
For two $A_\infty$-categories $\cC$ and $\cC'$, 
an $A_\infty$-functor $\cF:\cC\raw\cC'$ 
consists of a map between the objects $\f:\Ob(\cC)\raw\Ob(\cC')$ and a collection of linear maps 
$\cF:=\{\f_n:\Hom(a_1,a_2)\otimes\Hom(a_2,a_3)\otimes
\cdots\otimes\Hom(a_n,a_{n+1})\raw\Hom(\f(a_1),\f(a_{n+1}))\}_{\substack{a_i\in\Ob(\cC)\\n\ge 1}}$
satisfying the following conditions
\begin{equation}
 \begin{split}
&\sum_{k_1+\cdots+k_i=n}m'_i
\(\f_{k_1}\otimes\cdots\otimes\f_{k_i}\)
=\sum_{k+l=n+1}\sum_{j=0}^{k-1}
\f_k(\1^{\otimes j}\otimes m_l\otimes\1^{\otimes (n-j-l)})\ 
\end{split}
\label{amorphism}
\end{equation}
on $\Hom(a_1,a_2)\otimes\cdots\otimes\Hom(a_n,a_{n+1})$. 
Furthermore, for two cyclic $A_\infty$-categories $\cC$ and $\cC'$ 
with their inner products $\omega$ and $\omega'$, respectively, 
we call an $A_\infty$-functor $\cF:\cC\raw\cC'$ cyclic when 
\begin{equation}
 \omega'(\f_1(e_{ab}),\f_1(e_{ba}))=\omega(e_{ab},e_{ba})\ ,
 \label{omegacF1}
\end{equation}
and for fixed $n\ge 3$, 
\begin{equation}
 \sum_{k,l\ge 1,\ k+l=n}
\omega'(\f_k(e_{12},\cdots,e_{k(k+1)}),\f_l(e_{(k+1)(k+2)},\cdots,e_{n(n+1)}))=0\ 
 \label{omegacF2}
\end{equation}
holds. 
 \label{defn:Ainftyfunc}
\end{defn}
An $A_\infty$-functor $\cF:\cC\raw\cC'$ induces an $A_\infty$-morphism 
$\cF:(\cH,\m)\raw (\cH',\m')$, and a cyclic $A_\infty$-functor $\cF:\cC\raw\cC'$ 
induces a cyclic $A_\infty$-morphism $\cF:(\cH,\omega,\m)\raw (\cH',\omega',\m')$. 
\begin{defn}[Homotopy]
An $A_\infty$-functor $\cC\raw \cC'$ 
is called {\it homotopy} if $\f:\Ob(\cC)\raw\Ob(\cC')$ is a bijection and 
$\f_1:\Hom(a,b)\raw\Hom(\f(a),\f(b))$ induces an isomorphism on the cohomologies 
for each $a$ and $b$. 
Also, it is called {\it fully faithful} if 
$\f_1:\Hom(a,b)\raw\Hom(\f(a),\f(b))$ is an isomorphism. 
Alternatively, we call two cyclic $A_\infty$-categories $\cC$ and $\cC'$ are homotopic 
if there exists a cyclic $A_\infty$-functor $\cC\raw\cC'$ which defines a homotopy 
as $A_\infty$-categories. 
 \label{defn:Ainftyhomo}
\end{defn}
A homotopy and a fully faithful $A_\infty$-functor correspond to 
an $A_\infty$-quasi-isomorphism and an $A_\infty$-isomorphism, respectively, 
in the terminology of $A_\infty$-algebras. 
\begin{defn}[Original definition]
There exists another definition of an $A_\infty$-category $\cC$ 
which is related to the one in Definition \ref{defn:Ainfty} above. 
It consists of a set of objects $\Ob(\cC)$, a graded vector space $V_{ab}$ 
for each objects $a,b\in\Ob(\cC)$ and a collection of multilinear maps 
$\m:=\{m_n:V_{a_1a_2}\otimes\cdots\otimes V_{a_na_{n+1}}\raw V_{a_1a_{n+1}}\}$ 
of degree $(2-n)$ satisfying 
\begin{equation}
 \begin{split}
&0=\sum_{k+l=n+1}\sum_{j=0}^{k-1}
(-1)^\epsilon \\
&\ m_k(v_{12},\cdots,v_{j(j+1)},m_l(v_{(j+1)(j+2)},\cdots,v_{(j+l)(j+l+1)}),
 v_{(j+l+1)(j+l+2)},\cdots,v_{n(n+1)})\ ,
 \end{split}
 \label{Ainfty2}
\end{equation}
where $\epsilon=(j+1)(l+1)+l(|v_{12}|+\cdots+|v_{j(j+1)}|)$. 
On this $A_\infty$-category $\cC$, a cyclic structure is given %$(V,\m)$
by a nondegenerate symmetric bilinear map 
$\eta:V_{ab}\otimes V_{ba}\raw\C$ of fixed degree $|\eta|\in\Z$ 
satisfying 
\begin{equation}
  \eta(m_n(v_{12},\cdots,v_{n(n+1)}),v_{(n+1)1})
 =(-1)^{n+(|v_{23}|+\cdots +|v_{(n+1)1}|)|v_{12}|}
 \eta(m_n(v_{23},\cdots,v_{(n+1)1}),v_{12})\ ,
\end{equation}
for each $n\ge 1$. 
 \label{defn:Ainfty2}
\end{defn}
Again, for $V:=\oplus_{a,b}V_{ab}$, $(V,\m)$ forms an $A_\infty$-algebra \cite{Sta1,Sta11}, 
and $(V,\eta,\m)$ forms a cyclic $A_\infty$-algebra (see \cite{Poli2}). 
As a cyclic DGA is an example of a cyclic $A_\infty$-algebra, 
a cyclic DG category is an example of a cyclic $A_\infty$-category 
in Definition \ref{defn:Ainfty2} 
with $\eta=\eta$, $m_1=d$, $m_2=\varphi$ and $m_3=m_4=\cdots=0$. 
The definition of a minimal $A_\infty$-category is just the same, an 
$A_\infty$-category with $m_1=0$. 
One can see that, for a minimal $A_\infty$-category $\cC$ of the original definition, 
the $A_\infty$ relations (\ref{Ainftyrel}) for $n=3$ reduces to the associativity 
condition of the composition of morphisms $m_2$. 
Thus, $(\cH,m_2)$ forms a category in a usual sense.

Anyway, both definitions of (cyclic) $A_\infty$-categories are equivalent by the lemma below. 
\begin{lem}
For a graded vector space $V=\oplus_k V^k$ with $k$ the grading, let  
\begin{equation*}
 s: V^k\raw V^{k-1}[1]=:\cH^{k-1}\ ,\qquad v_{ab}\mapsto e_{ab}
\end{equation*}
be a degree shifting operator called a suspension, 
where the indices $a,b,\cdots$ are preserved by this operation $s$. 
Then two cyclic $A_\infty$-categories in Definition \ref{defn:Ainfty} 
and Definition \ref{defn:Ainfty2} are compatible with each other 
through the suspension $s$. 
 \label{lem:sus}
\end{lem}
\pf
The equivalence of the two $A_\infty$-categories is given in
\cite{GJ}. Let us distinguish the $A_\infty$-structures in two notations by 
$\m^\cH$ and $\m^V$. A relation between 
the multilinear maps is given by 
$m^\cH_n=(-1)^{\sum_{i=1}^{n-1}(n-i)}s m^V_n((s^{-1})^{\otimes n})$ 
or more explicitly 
\begin{equation}
 \begin{split}
 m^\cH_n(e_{12},\cdots,e_{n(n+1)})
&=(-1)^{\sum_{i=1}^{n-1}(n-i)e_{i(i+1)}}
s m^V_n(s^{-1}(e_{12}),\cdots,s^{-1}(e_{n(n+1)}))\\
&= (-1)^{\sum_{i=1}^{n-1}(n-i)e_{i(i+1)}}
s m^V_n(v_{12},\cdots,v_{n(n+1)})\ . 
 \end{split}
\end{equation}
A relation between the two cyclic structures are also given by 
$\omega=\eta(s^{-1}\ ,s^{-1}\ )$, or 
\begin{equation}
 \omega(e,e')=(-1)^e\eta(s^{-1}(e) ,s^{-1}(e'))\ .
\end{equation}
\qed
\begin{rem}
Through this relation, the definitions of minimal $A_\infty$-categories 
(Definition \ref{defn:minimal}), cyclic $A_\infty$-functor 
(Definition \ref{defn:Ainftyfunc}), and homotopies (Definition \ref{defn:Ainftyhomo}) 
can be translated into those for cyclic $A_\infty$-categories of Definition \ref{defn:Ainfty2}. 
 \label{rem:sus}
\end{rem}
For an $A_\infty$-category, the original definition (Definition \ref{defn:Ainfty2}) is natural 
from a homotopical point of view, where 
$m_2$ is a usual degree zero product, $m_3$ is a homotopy between the violation of 
associativity of $m_2$, and so on \cite{Sta1,Sta11}. 
However, Definition \ref{defn:Ainfty} is simpler in sign. 
Therefore, for a cyclic $A_\infty$-category or a cyclic DG category, 
we take Definition \ref{defn:Ainfty2} when we concentrate on the 
structure $m_2$ only, whereas in subsection \ref{ssec:main} 
we use Definition \ref{defn:Ainfty} where higher multilinear maps are
considered in general. 
\begin{defn}[Fukaya's $A_\infty$-category on (non)commutative two-tori]
For a fixed $\theta\in\R$, 
let us consider a two-torus $\Th^2$ whose covering space is $\wt{\Th^2}\simeq\R^2$ 
with coordinates $(x_1,x_2)\in\R^2$. We have $\pi_{12}\wt{\Th^2}=\Th^2$, 
$\pi_{12}:=\pi_1\pi_2=\pi_2\pi_1$, where $\pi_1$ and $\pi_2$ is the projections 
associated with the identifications $x_1\sim x_1+1$ and $x_2\sim x_2+1$, respectively.  
An object $a$ is a geodesic cycle $\pi_{12}(L_a)\in\Th^2$, 
\begin{equation}
 L_a : q_a x_2=p_a x_1+\alpha_a\ ,\qquad \alpha_a\in\R\ ,
\end{equation}
where $p_a$ and $q_a$ are relatively prime integers satisfying $q_a+p_a\theta>0$, 
with a trivial line bundle equipped with a flat connection 
\begin{equation}
 \nabla_{a,1}=\fpartial{x_1}-2\pi i\beta_a\ ,\qquad \beta_a\in\R \ .
\end{equation}
For each object $a$ we assign a number 
\begin{equation}
 \mu_a:=\frac{p_a}{q_a+p_a\theta}\ , 
\end{equation}
and for any two objects $a$ and $b$, we define 
\begin{equation}
 \mu_{ab}:=(\mu_b-\mu_a)(q_a+p_a\theta)^2
 =\frac{p_{ab}}{q_{ab}+p_{ab}\theta_a}\ .
\end{equation}
Note that by $SL(2,\R)$ translation 
\begin{equation*}
  \bp x_1 \\ x_2 \ep \raw 
  \bp x_1^\theta \\ x_2^\theta \ep=
\bp 1 & \theta \\ 0 & 1\ep 
   \bp x_1 \\ x_2 \ep\ ,
\end{equation*}
a line $x_1+\theta x_2=0$ becomes a vertical line $x_1^\theta=0$, 
$\mu_a$ is the slope of $L_a$ after the translation, 
and $\mu_{ab}$ is $(q_a+p_a\theta)^2$ times the difference of the slope. 
The space of morphisms $\Hom$ is a 
graded vector space. We denote by $k$ the grading and 
write $V^F_{ab}=\oplus_k V_{ab}^{F,k}$, $V^{F,k}_{ab}:=\Hom^k(a,b)$. 
In $\Th^2$ there exists $|p_{ab}|$ (transversal) intersection
points of $\pi_{12}(L_a)$ and $\pi_{12}(L_b)$ if $\mu_a\ne\mu_b$. 
These points are determined by projecting the intersection points of 
\begin{equation*}
 L_a : q_a x_2=p_a x_1+\alpha_a\ ,\qquad 
 L'_b : q_b\(x_2-p_a\frac{q_{ab}j}{p_{ab}}\)
 =p_b\(x_1-q_a\frac{q_{ab}j}{p_{ab}}\)+\alpha_b
\end{equation*}
by $\pi_{12}:\wt{\Th^2}\raw\Th^2$. 
We assign them by $v_{ab}^j$, $j=0,1,\cdots,|p_{ab}|-1$. 
We identify $\{v_{ab}^j\}$ with the basis of $V^F_{ab}$. 
The grading of $v_{ab}^j$ is 
$0$ if $\mu_{ab}>0$ and $1$ if $\mu_{ab}<0$. 
We denote the place of $v_{ab}^j$ by $(x_1,x_2)(v_{ab}^j)$. 
Note that 
\begin{equation*}
 (x_1,x_2)(v_{ba}^{-q_{ab}j})=(x_1,x_2)(v_{ab}^j)
\end{equation*}
holds. In case $\mu_{ab}=0$ (nontransversal case), 
$L_a$ and $L_b$ does not intersect with each other for $a\ne b$ and 
they coincide with each other if $a=b$. 
According to this, 
we formally introduce bases of morphisms $\1_a\in V^{F,0}_{aa}$ and 
${\bar \1}_a\in V^{F,1}_{aa}$. 
They are uniformly denoted in general expression by $v_{ab}^j$ with $a=b$ and $j=0$; 
for $a=b$, if $v_{ab}^0=\1_a$ then $v_{ba}^0={\bar \1}_a$, 
and if $v_{ab}^0={\bar \1_a}$ then $v_{ba}^0=\1_a$. 
An $A_\infty$ structure $\m^F$ is defined by 
$m^F_1=0$ and higher compositions $m^F_n$, $n\ge 2$ of degree $(2-n)$, 
which are given by 
\begin{equation}
 \begin{split}
& m^F_n(v^{j_{12}}_{a_1a_2},\cdots,v^{j_{n(n+1)}}_{a_na_{n+1}})=
 c_{a_1\cdots a_{n+1}}^{j_{12}\cdots j_{(n+1)1}}\cdot
 v^{j_{1(n+1)}}_{a_1a_{n+1}}\ ,\\
&c_{a_1\cdots a_{n+1}}^{j_{12}\cdots j_{(n+1)1}}
 =\sum_{\ti{v}\in I_{a_1\cdots a_{n+1}}^{j_{12}\cdots j_{(n+1)}}}
 sign(x_1^\theta(\ti{v}_{a_1a_2}^{j_{12}})
-x_1^\theta(\ti{v}_{a_{n+1}a_1}^{j_{(n+1)1}}))
 \exp{\(2\pi i\rhoh A(\ti{v})\)}\exp{(2\pi i\int\beta(\ti{v}))}\ 
 \end{split}
 \label{fukaya}
\end{equation}
if $\mu_{a_ia_{(i+1)}}\ne 0$ for any $i$. 
Here $I_{a_1\cdots a_{n+1}}^{j_{12}\cdots j_{(n+1)1}}$ 
is the subset of 
\begin{equation*}
\{\ti{v}=(\ti{v}^{j_{12}}_{a_1a_2},\cdots,\ti{v}^{j_{n(n+1)}}_{a_na_{n+1}},
\ti{v}^{j_{(n+1)1}}_{a_{n+1}a_1})\in 
(\pi_{12}^{-1}({v}^{j_{12}}_{a_1a_2}),\cdots,\pi_{12}^{-1}({v}^{j_{n(n+1)}}_{a_na_{n+1}})),
\pi_{12}^{-1}(v^{j_{(n+1)1}}_{a_{n+1}a_1})\}
\end{equation*}
satisfying the following conditions. 
\begin{itemize}
 \item $j_{(n+1)1}=-q_{a_{n+1}a_1}j_{1(n+1)}$. 
 \item Interval $(\ti{v}^{j_{(i-1)i}}_{a_{i-1}a_i},\ti{v}^{j_{i(i+1)}}_{a_ia_{i+1}})$ is included in 
 $\pi_{12}^{-1}\pi_{12}(L_{a_i})$ for each $i=1,\cdots,n+1$, 
 where $\ti{v}_{a_0a_1}^{j_{01}}=
\ti{v}_{a_{n+1}a_1}^{j_{(n+1)1}}=\ti{v}_{a_{n+1}a_{n+2}}^{j_{(n+1)(n+2)}}$. 
 \item $\ti{v}$ forms a convex. 
 \item Modulo an equivalence relation; 
 $\ti{v}$ and $\ti{v}'$ are identified with each other, if they coincides with a 
 parallel transformation on $\wt{\Th^2}$. We can remove this equivalence relation by 
 fixing a vertex, for instance, $\ti{v}^{j_{12}}_{a_1a_2}=v^{j_{12}}_{a_1a_2}\in \wt{\Th^2}$ 
(more precisely, the image of $v^{j_{12}}_{a_1a_2}$ by a natural inclusion 
$\Th^2\hraw\wt{\Th^2}$). 
\end{itemize}
$A(\ti{v})$ is then the area of the convex $(n+1)$-gon, 
and $\int\beta(\ti{v})$ is given by 
\begin{equation}
 \sum_{i=1}^{n+1} 
 (x_1(\ti{v}_{a_ia_{i+1}}^{j_{i(i+1)}})
 -x_1(\ti{v}_{a_{i-1}a_i}^{j_{(i-1)i}}))\beta_i \ ,
\end{equation}
where 
$x_1(\ti{v}_{a_ia_{i+1}}^{j_{i(i+1)}})$ is the $x_1$-coordinate 
of the point $\ti{v}_{a_ia_{i+1}}^{j_{i(i+1)}}$ on $\wt{\T^2}$. 
If at least one of $v_{a_ia_{i+1}}$, $i=1,\cdots,n+1$ is 
$\1_a$ in eq.(\ref{fukaya}), 
$m^F_n$ vanishes for $n\ge 3$ and only when $n=2$, 
by regarding the corresponding area of the 
triangle is zero, we set $c_{a_1a_2a_3}^{j_{12}j_{23}j_{31}}=1$. 
This indicates that $\1_a$ for each $a$ is the unit in this 
$A_\infty$-category. 
Alternatively, if at least one of $v_{a_ia_{i+1}}$ is 
${\bar \1}_a$, we set $c_{a_1\cdots a_{n+1}}^{j_{12}\cdots j_{(n+1)1}}=0$ 
for $n\ge 3$, and in case $n=2$ 
the rest two basis must be $\1_a$ so 
$c_{a_1a_2a_3}^{j_{12}j_{23}j_{31}}=1$.
\footnote{For the higher products $m^F_n$, $n\ge 3$ 
including ${\bar \1}_a$, the definition here is a tentative one. 
This in fact satisfies the $A_\infty$-relation (\ref{Ainfty2}), but 
there exist other choices. 
These higher products including $\1_a$ or ${\bar \1}_a$ 
should be defined from 
Floer homology in the situation that 
some of lines are nontransversal (coincides with each other). } 
One can define a degree minus one nondegenerate symmetric 
inner product $\eta^F:V^F_{ab}\otimes V^F_{ba}\raw\C$ by 
\begin{equation}
 \eta^F(v_{ab}^j,v_{ba}^{j'})=\delm{p_{ab}}_{-q_{ab}j}^{j'}\ ,
 \label{delm}
\end{equation}
where by $\delm{p_{ab}}$ we indicate the Kronecker's delta of mod $|p_{ab}|$, that is, 
$\delm{p_{ab}}_j^{j'}=1$ when $(j'-j)\in p_{ab}\Z$ and zero otherwise. 
Then $(V^F,\eta^F,\m^F)$ forms a cyclic $A_\infty$-category in Definition 
\ref{defn:Ainfty2}. Since $m_1=0$, this is a minimal cyclic $A_\infty$-category. 
One can see that $(V^F,\eta^F,\m^F)$ is essentially independent of 
the choice of $r_a$ and $s_a$. 
Though $q_{ab}=-s_ap_b+r_aq_b$,  
the ambiguity of the choice is absorbed into the permutations of 
$\{j=0,1,\cdots,|p_{ab}|-1\}$. 
 \label{defn:fukaya}
\end{defn}
The fact that only convex $(n+1)$-gons are `counted' is equivalent to 
the fact that $c_{a_1\cdots a_{n+1}}^{j_{12}\cdots j_{(n+1)1}}$ is 
nonzero only when 
$\sum_{i=1}^{n+1}\deg(v_{a_ia_{i+1}}^{j_{i(i+1)}})=-2+(n+1)$. 
The $A_\infty$-relation follows from concentrating on a polygon which 
has one nonconvex vertex. 
There exist two ways to divide the polygon into two convex polygons. 
The corresponding terms then appear with opposite sign and 
cancel each other in the $A_\infty$-relation. 
\begin{conj}[Homological mirror symmetry for noncommutative two-tori]
Let $\cC^A$ be a cyclic $A_\infty$-category in Definition \ref{defn:Ainfty} given by 
Fukaya's $A_\infty$-category $(V^F,\eta^F,\m^F)$ in Definition \ref{defn:fukaya} 
through the suspension in Lemma \ref{lem:sus}, and $\cC^B$ a cyclic $A_\infty$-category 
in Definition \ref{defn:Ainfty} given by the cyclic DG category $(V,\eta,d,\varphi)$ 
in Proposition \ref{prop:cDGcat} through the suspension. 
Then, two cyclic $A_\infty$-categories $\cC^A$, $\cC^B$ are homotopic to each other. 
 \label{conj:Ainfty}
\end{conj}
It is known that any $A_\infty$-category is homotopic to a minimal $A_\infty$-category. 
Such a homotopy algebraic property of $A_\infty$-categories is quite the same as that 
of $A_\infty$-algebras, for which 
this fact is shown by Kadeishvili \cite{kadei1} and called the 
minimal model theorem. 
A stronger version, decomposition theorem, also holds 
including the cyclic case \cite{thesis}. 
Thus, a standard way to examine the equivalence of two cyclic $A_\infty$-categories 
is to construct the corresponding minimal model of each cyclic $A_\infty$-categories. 
In this situation, a homotopy between the two minimal models is, if there exists, 
a fully faithful cyclic $A_\infty$-functor, which just corresponds to the 
ambiguity of minimal models. 
In our Conjecture \ref{conj:Ainfty}, since $(V^F,\eta^F,\m^F)$ is originally minimal, 
one may construct a minimal model of $(V,\eta,d,\varphi)$. 
As presented later in subsection \ref{ssec:main}, 
there is a way to construct a minimal model \cite{mer,KoSo} 
based on the reduction on the cohomology space with respect to $d=m_1$. 
The minimal model of $(V,\eta,d,\varphi)$ is then, 
by employing the Hodge-Kodaira decomposition $dd^{-1}+d^{-1}d+P=\1$ (Definition \ref{defn:HK}), 
a graded vector space $PV$ equipped with inner product $\eta|_{PV}$, product $P\varphi$, and higher 
products, where $|_{PV}$ indicates the restriction onto $PV$. 
On the other hand, 
by forgetting the higher products of a minimal (cyclic) $A_\infty$-category, 
one gets a (cyclic) graded category. 
If two cyclic $A_\infty$-categories are homotopy equivalent, 
the corresponding two cyclic graded categories must be equivalent. 
Here we show this. 
\begin{thm}
On (non)commutative two-tori, a cyclic graded category $(V^F,\eta^F,m_2^F)$ of 
the Fukaya's $A_\infty$-category is equivalent to a cyclic graded category 
$(PV,\eta|_{PV},P\varphi)$ of holomorphic vector bundles. 
Namely, there exists an isomorphism $\f_1:V_{ab}^{F,k}\raw PV_{ab}^k$ for any $a, b$ 
which is compatible with both the product structures and the cyclic structures 
\begin{equation}
\f_1 m_2^F= P\varphi(\f_1\otimes\f_1)\ ,\qquad 
\f_1\eta^F=\eta|_{PV}(\f_1\otimes\f_1)\ .
 \label{cycf}
\end{equation} 
 \label{thm:cat}
\end{thm}
\pf 
Since we already identify objects in both sides with each other 
and use the same notation $a,b,c,\cdots$, 
the map between the objects is just the identity map, 
For the morphisms, a map is defined by 
\begin{equation}
 \f_1 : v_{ab}^j \mapsto |0,0\ra_{ab}^j 
 \label{idf1}
\end{equation}
for degree zero morphisms and 
\begin{equation}
 \f_1 : v_{ab}^j \mapsto N_{ab}(0)|0,1\ra_{ab}^j 
 \label{idf2}
\end{equation}
for degree one morphisms. 
We shall denote the morphism corresponding to $v_{ab}^j$ also by the same notation 
$v_{ab}^j\in PV$. 
The compatibility of $\f_1$ with respect to the cyclicity in eq.(\ref{cycf}) is easily checked 
directly. 
Now we show that the tensor product on the cohomology of the cyclic DG
category $P\varphi$ coincides with the $m^F_2$ 
in the Fukaya's $A_\infty$-category side. 
The tensor product $\varphi(v^{j_{ab}}_{ab}\otimes v^{j_{bc}}_{bc})$ was 
computed in the case $\mu_{ab}>0$ and $\mu_{bc}>0$ 
in \cite{Stensor, foliation, PoSc}, where 
it was observed that 
the holomorphic vectors (elements of $H^0(V)$) 
close with respect to the tensor product. 
In general, the tensor product restricted on the cohomologies 
is given by 
\begin{equation*}
 P\varphi(v^{j_{ab}}_{ab}\otimes v^{j_{bc}}_{bc})
 =\sum_{n=0}^\infty P |n,k\ra_{ac}^j N_{ac}(n)
\eta(\varphi(v_{ab}^{j_{ab}}\otimes v_{bc}^{j_{bc}}),
 |n,1-k\ra_{ca}^{(-q_{ac}j)})\ , 
\end{equation*}
where $k$ is the appropriate degree (zero or one). Since $P|n,k\ra_{ac}^j=|0,k\ra_{ac}^j$, 
it is expressed as 
\begin{equation*}
 P\varphi(v^{j_{ab}}_{ab}\otimes v^{j_{bc}}_{bc})
 =c_{abc}^{j_{ab}j_{bc}(-q_{ac}j)}\cdot v_{ac}^j\ ,\qquad
 \ c_{abc}^{j_{ab}j_{bc}j_{ca}}:=
 \eta(\varphi(v_{ab}^{j_{ab}}\otimes v_{bc}^{j_{bc}}),v_{ca}^{j_{ca}})\ .
\end{equation*}
The definition of cyclic DG category guarantees that 
this structure constant $c_{abc}^{j_{ab}j_{bc}j_{ca}}\in\C$ is 
invariant with respect to the cyclic permutation 
$(a,b,c;j_{ab},j_{bc},j_{ca})\raw (b,c,a;j_{bc},j_{ca},j_{ab})$. 
Therefore, to see the coincidence of this tensor product with 
the Fukaya category sides, it is enough to 
check the coincidence of $c_{abc}^{j_{ab}j_{bc}j_{ca}}$ in both sides 
in the case $\deg(v_{ab}^{j_{ab}})=\deg(v_{bc}^{j_{bc}})=0$, hence 
$\deg(v_{ca}^j)=1$. 
There are the following four cases. 
\begin{equation}
 \begin{array}{lcccc}
 \mu_{ab}\ & =0, & =0, & >0, & >0, \\
 \mu_{bc}\ & =0, & >0, & =0, & >0, \\
 \mu_{ca}\ & =0, & <0, & <0, & <0, 
 \end{array}\quad .
\end{equation}
The first three cases are trivial. 
When $c_{abc}^{j_{ab}j_{bc}j_{ca}}$ includes at least one element of
$\mu=0$, it reduces to the inner product and then 
$c_{abc}^{j_{ab}j_{bc}j_{ca}}=1$. 
Let us compare $c_{abc}^{j_{ab}j_{bc}j_{ca}}$ in the last nontrivial case. 
For the tensor product side, by a direct calculation we get 
\begin{align}
 c_{abc}^{j_{ab}j_{bc}j_{ca}}
 &=\sum_{u\in\Z}\delm{p_{ab}}^{j_{ab}}_{-r_{ab}u-r_{ac}j_{ca}}\delm{p_{bc}}^{j_{bc}}_u
\exp{\(\frac{2\pi i}{-p_{ab}p_{bc}p_{ca}}
\(\frac{\rhoh}{2}
(l_{abc}-\alpha_{abc})^2-\beta_{abc}(l_{abc}-\alpha_{abc})\)\)}\ ,
\label{strconst}\\
 &l_{abc}(u,j_{ab},j_{bc},j_{ca}):=p_{ac}u+p_{bc}r_{ac}j_{ca}\ ,
 \label{labc}\\ 
 &\alpha_{abc}
:=p_{ab}(q_c+p_c\theta)\alpha_{bc}-p_{bc}(q_a+p_a\theta)\alpha_{ab}
 =p_{bc}\alpha_a+p_{ca}\alpha_b+p_{ab}\alpha_c\ ,
 \label{alpha-abc}\\
 &\beta_{abc}
:=p_{ab}(q_c+p_c\theta)\beta_{bc}-p_{bc}(q_a+p_a\theta)\beta_{ab}
=p_{bc}\beta_a+p_{ca}\beta_b+p_{ab}\beta_c\ .
 \label{beta-abc}
\end{align}
Here $\delm{p_{ab}}$ and 
$\delm{p_{bc}}$ are the Kronecker's delta of mod $|p_{ab}|$ and $|p_{bc}|$, 
respectively, defined in eq.(\ref{delm}).

On the other hand, in the A-model side (Definition \ref{defn:fukaya}), 
the $A_\infty$-structure $m_2$ is given by the summation of triangles 
made from lines $\pi_{12}^{-1}\pi_{12}(L_a,L_b,L_c)$ on the covering space $\wt{\Th^2}$. 
Let us set the following lines in $\wt{\Th^2}$; 
\begin{align}
 L_a:\ \ &q_a x_2=p_a x_1+\alpha_a\ ,\\
 L'_b:\ \ &q_b x_2=p_b x_1-q_{ab}j_{ab}+\alpha_b\ ,\\  
 L'_c:\ \ &q_c x_2=p_c x_1-q_{ac}j_{ab}
 -q_{bc}j_{bc}-p_{bc}m+\alpha_c\ .
\end{align}
$L'_b$ is the slope $p_b/q_b$ line 
which passes through 
$$
(x_1,x_2)=
\(q_a\frac{q_{ab}j_{ab}-\alpha_b}{p_{ab}},
p_a\frac{q_{ab}j_{ab}-\alpha_b}{p_{ab}}\)\ ,
$$ 
and $L'_c$ is as the slope $p_c/q_c$ line 
which passes through 
$$
(x_1,x_2)=\(q_a\frac{q_{ac}j_{ab}}{p_{ac}}+q_b\(m+\frac{q_{bc}j_{bc}-\alpha_c}{p_{bc}}\), 
p_a\frac{q_{ac}j_{ab}}{p_{ac}}+p_b\(m+\frac{q_{bc}j_{bc}-\alpha_c}{p_{bc}}\)
\)\ .
$$ 
We define $\ti{v}_{ab}^{j_{ab}}$ (resp. $\ti{v}_{bc}^{j_{bc}}$) by 
the intersection point of $(L_a,L'_b)$ (resp. $(L'_b,L'_c)$). 
\begin{align}
 &(x_1, x_2)(\ti{v}_{ab}^{j_{ab}})
=\(\frac{q_aq_{ab}j_{ab}-q_a\alpha_b+q_b\alpha_a}{p_{ab}}, 
\frac{p_aq_{ab}j_{ab}-p_a\alpha_b+p_b\alpha_a}{p_{ab}}\) 
\label{xy-eab}\\
 &(x_1,x_2)(\ti{v}_{bc}^{j_{bc}})
 =\(\frac{q_bp_{bc}m
+(q_bq_{ac}-q_cq_{ab})j_{ab}+q_bq_{bc}j_{bc}
-q_b\alpha_c+q_c\alpha_b}{p_{bc}}, \right.\\
&\hspace*{4.0cm} \left.
\frac{q_bp_{bc}m+(p_bq_{ac}-p_cq_{ab})j_{ab}+p_bq_{bc}j_{bc}
-p_b\alpha_c+p_c\alpha_b}{p_{bc}}\)\ .
 \label{xy-ebc}
\end{align}
Since the slopes of the three lines are given, 
the triangle is determined when the two points $v_{ab}^{j_{ab}}$ and 
$v_{bc}^{j_{bc}}$ are given. 
Namely, the triangle exists 
only when the intersection point of $(L_a,L'_c)$ 
\begin{equation*}
 (x_1,x_2)=\(\frac{q_a(p_{bc}m+q_{ac}j_{ab}+q_{bc}j_{bc})-q_a\alpha_c+q_c\alpha_a}
{p_{ac}}, 
\frac{p_ap_{bc}m+p_aq_{ac}j_{ab}+p_aq_{bc}j_{bc}-p_a\alpha_c+p_c\alpha_a}
{p_{ac}}\)\ 
\end{equation*}
coincides with 
\begin{equation*}
 (x_1,x_2)(\ti{v}_{ac}^{j_{ac}})
 =\(\frac{q_aq_{ac}j_{ac}-q_a\alpha_c+q_c\alpha_a}{p_{ac}}, 
\frac{p_aq_{ac}j_{ac}-p_a\alpha_c+p_c\alpha_a}{p_{ac}}\)\ ,
\quad -q_{ac}j_{ac}=j_{ca}\ ,
\end{equation*}
which is obtained by replacing $b,j_{bc}$ to $c,j_{ac}$ in eq.(\ref{xy-eab}). 
One can see that the condition is 
\begin{equation*}
 p_{ab}m+r_{ca}j_{ab}+q_{bc}j_{bc}+j_{ca} \in p_{ca}\Z\ .
\end{equation*}
The structure constant is then given by 
\begin{equation}
 c_{abc}^{j_{ab}j_{bc}j_{ca}}
 =\sum_{m\in\Z}\delm{p_{ca}}_{p_{bc}m+r_{ca}j_{ab}+q_{bc}j_{bc}+j_{ca}}^0
 \exp{\(2\pi i\rhoh\triangle_m\)}\exp{(2\pi i \int\beta_m)}\ ,
 \label{strconst2}
\end{equation}
where $\triangle_m$ is the area of the triangle made from 
$L_a$, $L'_b$ and $L'_c$ which is computed as 
\begin{equation}
 \triangle_m
=\ov{-2p_{ab}p_{bc}p_{ca}}
\(l'_{abc}(m,j_{ab},j_{bc},j_{ca})-\alpha_{abc}\)^2\ ,
 \label{tri}
\end{equation}
where $\alpha_{abc}$ is just that given in eq.(\ref{alpha-abc}) and 
\begin{equation}
 l'_{abc}(m,j_{ab},j_{bc},j_{ca}):=p_{ab}p_{bc}m+p_{ab}q_{bc}j_{bc}-p_{bc}j_{ab}\ .
 \label{l'}
\end{equation}
The effect of the holonomy $\beta$ is given by 
\begin{equation}
 \exp{(2\pi i \int\beta_m)}:=\exp{\(-2\pi i
 \beta_{abc}\(l'_{abc}(m,j_{ab},j_{bc},j_{ca})-\alpha_{abc}\)\)}\ ,
\end{equation}
where $\beta_{abc}$ is also that given in eq.(\ref{beta-abc}).

The rest of the proof is to show the coincidence between 
$l_{abc}$ and $l'_{abc}$  
under the Kronecker's delta's 
in eq.(\ref{strconst}) and eq.(\ref{strconst2}). 
First, 
$\delm{p_{ab}}^{j_{ab}}_{-r_{ab}u-r_{ac}j_{ca}}
=\delm{p_{ab}}^{-q_{ab}j_{ab}}_{u+q_{ab}r_{ac}j_{ca}}$ 
can be deleted by putting 
\begin{equation}
 u=-q_{ab}j_{ab}-q_{ab}r_{ac}j_{ca}+p_{ab}(v+s_{ac}j_{ca})
=p_{ab}v-q_{ab}j_{ab}-r_{bc}j_{ca}
 \label{u-v}
\end{equation}
for $v\in\Z$. The term $s_{ac}j_{ca}$ is included for convenience. 
Substituting this $u$ into $l_{abc}$ one gets 
\begin{equation}
 l_{abc}(u=p_{ab}v-q_{ab}j_{ab}-r_{bc}j_{ca})
 =-p_{ca}p_{ab}v+p_{ca}q_{ab}j_{ab}-p_{ab}j_{ca}\ .
 \label{l2}
\end{equation}
On the other hand, combining eq.(\ref{u-v}) with $\delm{p_{bc}}^{j_{bc}}_u$ 
one gets a constraint for $v$ as follows; 
\begin{equation}
 \begin{split}
 0=&q_{bc}\(j_{bc}-(p_{ab}v-q_{ab}j_{ab}-r_{bc}j_{ca})\)+p_{bc}m\ \\
  =&p_{bc}m'+r_{ca}j_{ab}+q_{bc}j_{bc}+j_{ca}+p_{ca}v
 \end{split}
 \label{v-constraint}
\end{equation}
for $m\in\Z$ and $m'=m+r_{ab}v-s_{ab}j_{ab}-s_{bc}j_{ca}$. 
We denote by $I_{abc}^{j_{ab}j_{bc}j_{ca}}$ 
the set of $v$ satisfies the constraint above; 
\begin{equation}
 I_{abc}^{j_{ab}j_{bc}j_{ca}}:=\{v| \mbox{Exists}\ (v,m)\in \Z^2 s.t.\ 
 0=p_{bc}m+r_{ca}j_{ab}+q_{bc}j_{bc}+j_{ca}+p_{ca}v\}\ .
 \label{vm-set}
\end{equation}
Now, one can see that this constraint (\ref{vm-set}) is just the same
as $\delm{p_{ca}}_{p_{bc}m+r_{ca}j_{ab}+q_{bc}j_{bc}+j_{ca}}^0$ 
in eq.(\ref{strconst2}), and $l'_{abc}$ in eq.(\ref{l'}) is equal to 
$l_{abc}$ in eq.(\ref{l2}) under this constraint. 
Thus, we completed the coincidence of $c_{abc}^{j_{ab}j_{bc}j_{ca}}$ 
in (\ref{strconst}) and (\ref{strconst2}). 
 
Note that though the cyclicity of $c_{abc}^{j_{ab}j_{bc}j_{ca}}$ is guaranteed by
the cyclicity of $\V$, now it can be seen explicitly. 
$\alpha_{abc}$ and $\beta_{abc}$ are written just in manifest 
cyclic expression. Then $l_{abc}$ in eq.(\ref{l2}) and 
$l'_{abc}$ in eq.(\ref{l'}) are related by cyclic 
permutation of $abc$ and $j_{ab}j_{bc}j_{ca}$ 
(with a flop $v\lraw m$). Moreover, instead of eq.(\ref{v-constraint}), 
if one represent the constraint as 
\begin{equation}
  0=\(j_{bc}-(p_{ab}v-q_{ab}j_{ab}-r_{bc}j_{ca})\)+p_{bc}m\ ,
\end{equation}
one can see that this is also related to eq.(\ref{v-constraint}) by 
the cyclic permutation. 

Note that by the definition of $\alpha_{abc}$ (\ref{alpha-abc}), 
$\beta_{abc}$ (\ref{beta-abc}) and the expression of 
$l_{abc}^{j_{ab}j_{bc}j_{ca}}$ in (\ref{l2}), one can see that 
the structure constant $c_{abc}^{j_{ab}j_{bc}j_{ca}}$ depend only on 
$p_{ab},q_{ab}$ and 
is independent of $r_{ab}, s_{ab}$ in $\Hom(E_a,E_b)$. 
This fact justifies 
the correspondence of these two categories.

\qed

For commutative two-tori, 
the categorical mirror symmetry is given by Polishchuk-Zaslow \cite{PZ}, 
where a map $\f_1$ is given and shown to be compatible with the product structures on $H^0$ on both sides. 
Our result above, for $\theta=0$, 
gives an explicit and stronger version of it in the sense that we showed it on all the cohomologies 
$H^0$ and $H^1$ together with the cyclicity.

 \subsection{Higher products associated to polygons}
\label{ssec:main}

Recall that $\cC^B=(\cH,\omega^B,\m^B)$ is a cyclic $A_\infty$-category 
in Definition \ref{defn:Ainfty} given by the suspension (Lemma \ref{lem:sus}) 
of the cyclic DG category $(V,d,\varphi,\eta)$ in Proposition \ref{prop:cDGcat}. 
We relate $V$ and $\cH$ by the suspension $s$ as 
\begin{equation}
 \begin{array}{rccc}
 s : &V_{ab}^k & \raw &\cH_{ab}^{k-1}\\
     & |n,k\ra_{ab}^j   & \mapsto & |n,k-1)_{ab}^j\ .
 \end{array}
\end{equation}
The corresponding cyclic $A_\infty$-category is given explicitly as 
$m_1^B=d$ and 
\begin{equation}
 \begin{split}
& m^B_2(|n,k)_{ab}^j,|n',k')_{bc}^{j'})
 =(-1)^k\varphi(|n,k+1\ra_{ab}^j,|n',k'+1\ra_{bc}^{j'})\ ,\\
& \omega^B(|n,k)_{ab}^j,|n',k')_{ba}^{j'})
 =(-1)^k\eta(|n,k+1\ra_{ab}^j,|n',k'+1\ra_{ba}^{j'})\ .
 \end{split}
\end{equation}
We also define vertex map $\V:\cH_{ab}\otimes\cH_{bc}\otimes\cH_{ca}\raw\C$ by 
$\V:=\omega^B(m_2^B\otimes\1)$ for later convenience. We have 
\begin{equation}
 \V(|n,k)_{ab}^j,|n',k')_{bc}^{j'},|n'',k'')_{ca}^{j''})
 =(-1)^{k'+1}\eta(\varphi(|n,k+1\ra_{ab}^j,|n',k'+1\ra_{bc}^{j'}),
 |n'',k''+1\ra_{ca}^{j''})\ .
\end{equation}

Note that, in the shifted degree, the dimension of $H^k(\cH):=P\cH$ can be
nonzero only if $k$ is equal to minus one or zero. 

As stated previously, it is known that, for any $A_\infty$-algebra, there exists a minimal 
$A_\infty$-algebra which is $A_\infty$-quasi-isomorphic to the
original one \cite{kadei1}. 
An explicit way to construct a minimal 
$A_\infty$-structure $\ti\m^B=\{\ti{m}^B_2,\ti{m}^B_3,\cdots\}$ 
is given for instance in \cite{mer,KoSo} based on 
homological perturbation theory (cf.\cite{GS,GLS}). 
Since an $A_\infty$-category can be thought of as an $A_\infty$-algebra 
(Remark \ref{rem:cat-alg}), 
the story is quite the same for an $A_\infty$-category. 
We recall the way in concentrating on the case that the higher products 
of the original $A_\infty$-structures are absent; $m^B_3=m^B_4=\cdots=0$. 
For some related literatures in physics, see \cite{LazR,Tom,Laz,Dbcat}. 
First, for a given Hodge-Kodaira decomposition 
$dd^{-1}+d^{-1}d+P=1$ for $d=m_1^B$, 
construct a correction of degree zero (degree preserving) 
multilinear maps 
$\f_n:H(\cH_{a_1a_2})\otimes\cdots\otimes H(\cH_{a_na_{n+1}})
\raw\cH_{a_1a_{n+1}}$ inductively by 
\begin{equation}
 \f_n=\sum_{k,l\ge 1,\ k+l=n}d^{-1}
 m^B_2(\f_k\otimes\f_l)\ 
 \label{quasi-iso}
\end{equation}
with $\f_1=\iota$ and $\f_2=d^{-1}m^B_2$. 
The $A_\infty$ structure on the cohomology is then 
given by degree one multilinear maps 
$\ti{m}^B_n:H(\cH_{a_1a_2})\otimes\cdots\otimes H(\cH_{a_na_{n+1}})
\raw H(\cH_{a_1a_{n+1}})$ as follows 
\begin{equation}
 \ti{m}^B_n=\sum_{k,l\ge 1,\ k+l=n}P m^B_2(\f_k\otimes\f_l)\ 
 \label{minimal}
\end{equation}
for $\ti{m}^B_2= Pm^B_2$. 
One can also express this construction in terms of (tree level) 
Feynman graphs (see \cite{KoSo,infty,thesis}). 
Furthermore, when $d^{-1}$ satisfies 
\begin{equation}
 \omega^B(d^{-1}\ ,\ )=\pm\omega^B(\ ,d^{-1}\ )
 \label{cd-mincyc}
\end{equation}
the induced minimal $A_\infty$-structure 
$\ti\m^B=\{\ti{m}^B_n\}_{n\ge 2}$ has cyclicity with respect to $\omega|_{P\cH}$, 
the inner product restricted onto the cohomology \cite{infty,thesis}. 
The condition (\ref{cd-mincyc}) holds if we take for $d^{-1}$ the one 
given in Definition \ref{defn:HK}. 
Thus, $H(\cH,\omega,\m^B):=(P\cH,\omega |_{PV},\ti\m^B)$ 
forms a minimal cyclic $A_\infty$-category. 
In fact $\{\f_n\}_{n\ge 1}$ is a cyclic $A_\infty$-functor from 
the $A_\infty$-category constructed above to 
the cyclic DG category in Lemma \ref{prop:cDGcat}. 

Now, let us consider a vertex $\V$ 
\begin{equation}
 \V(f_{ab},f_{bc},f_{ca}):=\eta(m^B_2(f_{ab}\otimes f_{bc}),f_{ca})
\end{equation}
for $f_{ab}\in\cH_{ab}$, $f_{bc}\in\cH_{bc}$ and $f_{ca}\in\cH_{ca}$. 
The product $m_2$ is written by inverting vertices as 
\begin{equation} 
 m^B_2(f_{ab}\otimes f_{bc})=
\sum_{n}|n,-1+k)_{ac}^j N_{ab}(n)
 \V(f_{ab},f_{bc},|n,-k)_{ca}^{-q_{ac}j})\ ,
 \label{mv}
\end{equation}
where $N_{ab}(n)\in\R$ is the normalization in eq.(\ref{norma}). 
We need only trivalent vertices one of the edges is the leaf 
of $H^0(\cH_{ca})$ with $\mu_{ca}\le 0$. 
The reason is as follows. 
The vertex $\V$ is related to $m^B_2$ through eq.(\ref{mv}), 
and $m^B_2$ connects to either $d^{-1}$ as in eq.(\ref{quasi-iso}) or 
$P$ as in (\ref{minimal}). 
In the case of $d^{-1}m^B_2$, it does not vanish only if 
the image of $m^B_2$ has degree zero. This implies $k=1$ in
eq.(\ref{mv}), namely, the degree of elements 
which connect to internal edges are minus one. 
Elements of degree zero then contribute only to the leaves in terms of the tree graphs, 
that is, elements in $H^0(\cH)$. 
On the other hand, in eq.(\ref{mv}) when $k=1$, 
one of $f_{ab},f_{bc}$ has degree zero and another 
has degree minus one. 
Thus, for $d^{-1}m^B_2$ the corresponding vertex has an element 
in $H^0(\cH)$. 
As a result, eq.(\ref{quasi-iso}) can be reduced to 
\begin{equation*}
 \f_{n+1}=d^{-1}
 m^B_2(\f_n\otimes\f_1+\f_1\otimes\f_n)\ . 
\end{equation*}
In the case of $P$ acting on $m^B_2$ in eq.(\ref{mv}), 
again the vertex contains a degree zero element. 
If it is $f_{ab}$ or $f_{bc}$, the story is the same as above. 
If $|n,-k)_{ca}^{-q_{ac}j}$, 
only $|0,-k)_{ca}^{-q_{ac}j}$ survives because 
of the projection $P$.

Now we write down all the type of the vertices. 
By the cyclic permutation, let us set the elements of $H^0(\cH)$ 
as $|0,0)_{ca}^{j_{ca}}$ and consider vertex of the form 
\begin{equation}
 \V(|n,-1)_{ab}^{j_{ab}},|m,-1)_{bc}^{j_{bc}},|0,0)_{ca}^{j_{ca}}) \ .
\end{equation}
There are eight types of vertices as follows. 
\begin{equation}
 \begin{array}{lcccccccc}
          &_{(000)}&_{(><0)}&_{(<>0)}&_{(0><)}&_{(>0<)}
 &_{(>><)}&_{(><<)}&_{(<><)} \\
 \mu_{ab} & 0 & > & < & 0 & > & > & > & < \\
 \mu_{bc} & 0 & < & > & > & 0 & > & < & > \\
 \mu_{ca} & 0 & 0 & 0 & < & < & < & < & < \\
 \end{array}\ 
\end{equation}
where $0$, $>$, $<$ indicate the sign of $\mu_{**}$. 

In the first three cases (${(000)},{(><0)},{(<>0)}$), 
since $c=a$, the vertices just reduces to the inner product 
and we get 
\begin{equation}
 \V(|n,-1)_{ab}^{j_{ab}},|m,-1)_{ba}^{j_{ba}},|0,0)_a)
 =\ov{N_{ab}(n)}\delta_{n,m}\ .
\end{equation}
Next, let us consider the following type 
\begin{equation}
  \V(|n,-1)_{ab}^{j_{ab}},|0,-1)_{bc}^{j_{bc}},|0,0)_{ca}^{j_{ca}})\ 
\end{equation}
for $\mu_{ab}>0$ and $\mu_{bc}>0$. 
This is also expressed by using the Hermite polynomial 
as $|n,-1)_{ab}^j$ is ; 
\begin{equation}
 \begin{split}
& \V(|n,-1)_{ab}^{j_{ab}},|0,-1)_{bc}^{j_{bc}},|0,0)_{ca}^{j_{ca}})\ 
=\ov{N_{ac}(0))}
\frac{(2\pi T(p_{abc}))^{\frac{n}{2}}}{(q_b+p_b\theta)^n(p_{ca})^n}
\\
&\sum_{v\in I_{abc}^{j_{ab}j_{bc}j_{ca}}}
H_n\(\sqrt\frac{\pi T}{p_{abc}}
\lb_{abc}^{j_{ab}j_{bc}j_{ca}}(v)\)
\exp{\(\frac{2\pi i}{p_{abc}}
\(\frac{\rhoh}{2}(\lb_{abc}^{j_{ab}j_{bc}j_{ca}}(v))^2 
-\beta_{abc}\lb_{abc}^{j_{ab}j_{bc}j_{ca}}(v)\)\)}\ , 
 \end{split}\label{key-vertex}
\end{equation} 
where we put $\lb_{abc}^{j_{ab}j_{bc}j_{ca}}(v)
:=l_{abc}^{j_{ab}j_{bc}j_{ca}}(v)-\alpha_{abc}$ and 
$p_{abc}:=-p_{ab}p_{bc}p_{ca}\
(=-\mu_{ab}\mu_{bc}\mu_{ca})$. 
Since they are cyclic, 
except the coefficient, $H_n(\cdots)\exp{(\cdots)}$ 
in eq.(\ref{key-vertex}) is invariant with respect to the cyclic
permutation permutation $(a,b,c;j_{ab},j_{bc},j_{ca})\raw
(b,c,a;j_{bc},j_{ca},j_{ab})$. 
The other types are obtained as follows. First, 
using the Leibniz rule (Lemma \ref{lem:Leibniz}) we get 
${(>><)}$ type vertices 
\begin{equation}
 \V(|n,-1)_{ab}^{j_{ab}},|m,-1)_{bc}^{j_{bc}},|0,0)_{ca}^{j_{ca}})
 =(-1)^m \V(|n+m,-1)_{ab}^{j_{ab}},|0,-1)_{bc}^{j_{bc}},|0,0)_{ca}^{j_{ca}})\ .
 \label{cyclic-vertex}
\end{equation}
Here note that $\nabb^\dag|0,0)_{ca}^{j_{ca}}=0$. 
Next, replacing $\rhoh$ with $\rhoh^*$ in eq.(\ref{cyclic-vertex}) 
we get 
\begin{align}
&\V(|n,-1)_{ab}^{j_{ab}},|m,-1)_{bc}^{j_{bc}},|0,0)_{ca}^{j_{ca}})\nonumber\\ 
&\quad =-\V(|n,-1)_{ab}^{j_{ab}},|m,0)_{bc}^{j_{bc}},|0,-1)_{ca}^{j_{ca}}) 
 \label{reverse1} \\
&\quad =\V(|n,0)_{ab}^{j_{ab}},|m,-1)_{bc}^{j_{bc}},|0,-1)_{ca}^{j_{ca}})
 \label{reverse2}
\end{align}
for $\mu_{ab}<0$ and $\mu_{bc}<0$ (hence $\mu_{ca}>0$). 
More explicitly, 
by setting $m=0$ and permuting 
$(a,b,c;j_{ab},j_{bc},j_{ca})\raw (b,c,a;j_{bc},j_{ca},j_{ab})$ 
in eq.(\ref{reverse1}) we obtain 
\begin{equation}
 \begin{split}
& \V(|0,-1)_{ca}^{j_{ca}},|n,-1)_{ab}^{j_{ab}},|0,0)_{bc}^{j_{bc}})\ 
=\ov{N_{ba}(0)}
\frac{(-2\pi T(p_{abc}))^{\frac{n}{2}}}{(q_c+p_c\theta)^n(p_{ab})^n}
\\
&\sum_{v\in I_{abc}^{j_{ab}j_{bc}j_{ca}}}
H_n\(\sqrt\frac{-\pi T}{p_{abc}}
\lb_{abc}^{j_{ab}j_{bc}j_{ca}}(v)\)
\exp{\(\frac{2\pi i}{p_{abc}}
\(\frac{\rhoh^*}{2}(\lb_{abc}^{j_{ab}j_{bc}j_{ca}}(v))^2 
-\beta_{abc}\lb_{abc}^{j_{ab}j_{bc}j_{ca}}(v)\)\)}\ , 
 \end{split}\label{key-vertex1}
\end{equation}
and by setting $n=0$ and permuting 
$(a,b,c;j_{ab},j_{bc},j_{ca})\raw (c,a,b;j_{ca},j_{ab},j_{bc})$ 
in eq.(\ref{reverse2}) we get a similar result 
\begin{equation}
 \begin{split}
& \V(|m,-1)_{ab}^{j_{ab}},|0,-1)_{bc}^{j_{bc}},|0,0)_{ca}^{j_{ca}})\ 
=\ov{N_{cb}(0)}
\frac{(-2\pi T(p_{abc}))^{\frac{n}{2}}}{(q_a+p_a\theta)^n(p_{bc})^n}
\\
&\sum_{v\in I_{abc}^{j_{ab}j_{bc}j_{ca}}}
H_m\(\sqrt\frac{-\pi T}{p_{abc}}
\lb_{abc}^{j_{ab}j_{bc}j_{ca}}(v)\)
\exp{\(\frac{2\pi i}{p_{abc}}
\(\frac{\rhoh^*}{2}(\lb_{abc}^{j_{ab}j_{bc}j_{ca}}(v))^2 
-\beta_{abc}\lb_{abc}^{j_{ab}j_{bc}j_{ca}}(v)\)\)}\ .
 \end{split}\label{key-vertex2}
\end{equation}
On the other hand, for the ${(><<)}$ type vertex, 
\begin{equation}
 \V(|n,-1)_{ab}^{j_{ab}},|m,-1)_{bc}^{j_{bc}},|0,0)_{ca}^{j_{ca}})=
\frac{m!}{(m-n)!}([\nabb_{bc},\nabb_{bc}^\dag])^n
 \V(|0,-1)_{ab}^{j_{ab}},|m-n,-1)_{bc}^{j_{bc}},|0,0)_{ca}^{j_{ca}})
 \label{acyc-vertex1}
\end{equation}
holds for $m\ge n$ and 
$\V(|n,-1)_{ab}^{j_{ab}},|m,-1)_{bc}^{j_{bc}},|0,0)_{ca}^{j_{ca}})=0$ 
for $m<n$ due to the Leibniz rule. Similarly, for the ${(<><)}$ type
vertex we have 
\begin{equation}
 \V(|n,-1)_{ab}^{j_{ab}},|m,-1)_{bc}^{j_{bc}},|0,0)_{ca}^{j_{ca}})=
\frac{n!}{(n-m)!}(-[\nabb_{ab},\nabb_{ab}^\dag])^m
 \V(|n-m,-1)_{ab}^{j_{ab}},|0,-1)_{bc}^{j_{bc}},|0,0)_{ca}^{j_{ca}})
 \label{acyc-vertex2}
\end{equation}
for $n\ge m$ and 
$\V(|n,-1)_{ab}^{j_{ab}},|m,-1)_{bc}^{j_{bc}},|0,0)_{ca}^{j_{ca}})=0$ for $n<m$. 
The right hand sides of eq.(\ref{acyc-vertex1}) and
eq.(\ref{acyc-vertex2}) are just given by eq.(\ref{reverse1}) and 
eq.(\ref{reverse2}), respectively. 

The remaining vertices are those including $|n,k)_{ab}^{j_{ab}}$ with $\mu_{ab}=0$. 
They are also obtained by direct calculations. The ${(0><)}$ type vertex is
given by  
\begin{equation}
 \begin{split}
 &\V(|n,-1)_{aa'},|0,-1)_{a'b}^{j_{a'b}},|0,0)_{ba}^{j_{ba}})
 =\ov{N_{ab}(0)}
 \delm{p_{ab}}_{j_{ba}}^{-q_{ab}(j_{a'b}-n_2)}
 \exp{\(2\pi in_1\frac{j_{a'b} q_{ab}}{p_{ab}}\)} \\
 &\ \exp{\(\frac{\pi(q_a+p_a\theta)(q_b+p_b\theta)}{p_{ab}}
 \(-\ov{T}(n_1'+\rhoh n_2')^2+i\rhoh(n_2')^2
 +2i\(\alpha_{ab}n_1'+\beta_{a'b}n_2'+\alpha_{ab}\beta_{aa'}\)\)\)}\ ,
 \end{split}
\end{equation}
where $n_1':=\frac{n_1}{q_a+p_a\theta}-2\pi i\beta_{aa'}$ and 
$n_2':=\frac{n_2}{q_a+p_a\theta}-2\pi i\alpha_{aa'}$. 
Similarly, the ${(>0<)}$ type vertex is 
\begin{equation}
 \begin{split}
 &\V(|0,-1)_{ab}^{j_{ab}},|n,-1)_{bb'},|0,0)_{b'a}^{j_{b'a}})
 =\ov{N_{ab}(0)}
 \delm{p_{ab}}_{j_{b'a}}^{-q_{ab}(j_{ab}-n_2r_{ab})}
 \exp{\(2\pi in_1\(\frac{j_{ab}}{p_{ab}}-n_2\theta_b\)\)} \\
 &\ \exp{\(\frac{\pi(q_a+p_a\theta)(q_b+p_b\theta)}{p_{ab}}
 \(-\ov{T}(n_1'+\rhoh n_2')^2+i\rhoh (n_2')^2
 +2i\(\alpha_{ab'}n_1'+\beta_{ab}n_2'+\alpha_{ab'}\beta_{bb'}\)\)\)}\ ,
 \end{split}
\end{equation}
where $n_1':=\frac{n_1}{q_b+p_b\theta}-2\pi i\beta_{bb'}$ and 
$n_2':=\frac{n_2}{q_b+p_b\theta}-2\pi i\alpha_{bb'}$. 

We then get, due to the Leibniz rule, 
\begin{equation}
 \begin{split}
 & \V(|n,-1)_{aa'},|m,-1)_{a'b}^{j_{a'b}},|0,0)_{ba}^{j_{ba}})
 =\ov{(-(n_1'+\rhoh n_2'))^m}
 \V(|n,-1)_{aa'},|0,-1)_{a'b}^{j_{a'b}},|0,0)_{ba}^{j_{ba}}) \\
 & \V(|m,-1)_{ab},|n,-1)_{bb'}^{j_{bb'}},|0,0)_{b'a}^{j_{b'a}})
 =\ov{(-(n_1'+\rhoh n_2'))^m}
 \V(|0,-1)_{ab},|n,-1)_{bb'}^{j_{bb'}},|0,0)_{b'a}^{j_{b'a}})\ .
 \end{split}
\end{equation} 
Thus, we completed to determine all type of vertices and so 
the Feynman rules.

Lastly we end with giving an example of 
$\ti{m}^B_3$ with 
\begin{equation}
 \mu_d<\mu_c<\mu_a<\mu_b\ .
\end{equation}
In this case the basis of the cohomologies are 
$e_{ab}^{j_{ab}}=|0,-1)_{ab}^{j_{ab}}\in H^{-1}(\cH_{ab})$, 
$e_{bc}^{j_{bc}}=N_{bc}(0)|0,0)_{bc}^{j_{bc}}\in H^0(\cH_{bc})$, 
$e_{cd}^{j_{cd}}=N_{cd}(0)|0,0)_{cd}^{j_{cd}}\in H^0(\cH_{cd})$ and 
$e_{ad}^{j_{ad}}=N_{ad}(0)|0,0)_{ad}^{j_{ad}}\in H^0(\cH_{ad})$. 
Since $m^B_2(H^0\otimes H^0)$ vanishes by degree counting, we get 
\begin{equation}
 \begin{split}
 \ti{m}^B_3(e_{ab}^{j_{ab}},e_{bc}^{j_{bc}},e_{cd}^{j_{cd}})
 &=Pm_2(d^{-1}m_2(e_{ab}^{j_{ab}},e_{bc}^{j_{bc}}),e_{cd}^{j_{cd}})\\
 &=\V(e_{ab}^{j_{ab}},e_{bc}^{j_{bc}},|n,-1)_{ca}^{-q_{ac}j_{ac}})N_{ac}(n)
   \V(d^{-1}|n,0)_{ac}^{j_{ac}}, e_{cd}^{j_{cd}},|0,-1)_{da}^{-q_{ad}j_{ad}})  
\cdot e_{ad}^{j_{ad}}\ ,
 \end{split}
\end{equation}
where $d^{-1}|n,0)_{ac}^{j_{ac}}=|n-1,-1)_{ac}^{j_{ac}}$. 
By the Feynman rules obtained above, we have 
\begin{equation}
 \begin{split}
  &\V(e_{ab}^{j_{ab}},e_{bc}^{j_{bc}},|n,-1)_{ca}^{-q_{ac}j_{ac}})
  =-\V(|n,-1)_{ca}^{-q_{ac}j_{ac}},e_{ab}^{j_{ab}},e_{bc}^{j_{bc}})\\
  &\qquad =-\frac{\(2\pi T(p_{abc})\)^{\frac{n}{2}}}
{(q_a+p_a\theta)^n(p_{bc})^n} \ 
 \sum_{v\in I_{abc}^{j_{ab}j_{bc}(-q_{ac}j_{ac})}}
H_n\(\sqrt\frac{\pi T}{p_{abc}}
\lb_{abc}^{j_{ab}j_{bc}(-q_{ac}j_{ac})}(v)\) \\
& \hspace*{3.0cm}\cdot\exp{\(\frac{2\pi i}{p_{abc}}
\(\frac{\rhoh}{2}(\lb_{abc}^{j_{ab}j_{bc}(-q_{ac}j_{ac})}(v))^2 
-\beta_{abc}\lb_{abc}^{j_{ab}j_{bc}(-q_{ac}j_{ac})}(v)\)\)}
\ , \\
  &\V(|n-1,-1)_{ac}^{j_{ac}}, e_{cd}^{j_{cd}},|0,-1)_{da}^{-q_{ad}j_{ad}})
  =-\V(|n-1,-1)_{ac}^{j_{ac}}, N_{cd}(0)|0,-1)_{cd}^{j_{cd}},
  |0,0)_{da}^{-q_{ad}j_{ad}})  \\
  &=-\frac{N_{cd}(0)}{N_{ad}(0)}
 \frac{\(-2\pi T(p_{acd})\)^{\frac{n-1}{2}}}
{(q_c+p_c\theta)^{n-1}(p_{da})^{n-1}}\ 
\sum_{w\in I_{acd}^{j_{ac} j_{cd}(-q_{ad}j_{ad})}}
H_{n-1}\(\sqrt\frac{-\pi T}{p_{acd}}
\lb_{acd}^{j_{ac} j_{cd}(-q_{ad}j_{ad})}(w)\) \\
 &\hspace*{3.5cm}\cdot\exp{\(\frac{2\pi i}{p_{acd}}
\(\frac{\rhoh^*}{2}(\lb_{acd}^{j_{ac} j_{cd}(-q_{ad}j_{ad})}(w))^2 
-\beta_{acd}\lb_{acd}^{j_{ac} j_{cd}(-q_{ad}j_{ad})}(w)\)\)}\ .
 \end{split}
\end{equation}
Then we get 
\begin{equation}
 \begin{split}
& \ti{m}^B_3(e_{ab}^{j_{ab}},e_{bc}^{j_{bc}},e_{cd}^{j_{cd}})=
  \sum_{v\in I_{abc}^{j_{ab}j_{bc}(-q_{ac}j_{ac})}}
 \sum_{w\in I_{acd}^{j_{ac} j_{cd}(-q_{ad}j_{ad})}} \\
&\hspace*{0.5cm} \exp{\(\frac{2\pi i}{p_{abc}}
\(\frac{\rhoh}{2}(\lb_{abc}^{j_{ab}j_{bc}(-q_{ac}j_{ac})}(v))^2 
-\beta_{abc}\lb_{abc}^{j_{ab}j_{bc}(-q_{ac}j_{ac})}(v)\)\)} \\
&\hspace*{1.0cm}
\sum_{n=1}^\infty\ov{n!}\(\frac{p_{ab}p_{cd}}{p_{bc}p_{da}}\)^{\frac{n}{2}}
H_n\(\sqrt\frac{\pi T}{p_{abc}}
\lb_{abc}^{j_{ab}j_{bc}(-q_{ac}j_{ac})}(v)\)
H_{n-1}\(\sqrt\frac{-\pi T}{p_{acd}}
\lb_{acd}^{j_{ac} j_{cd}(-q_{ad}j_{ad})}(w)\) \\
&\hspace*{3.0cm} \exp{\(\frac{2\pi i}{p_{acd}}
\(\frac{\rhoh^*}{2}(\lb_{acd}^{j_{ac} j_{cd}(-q_{ad}j_{ad})}(w))^2 
-\beta_{acd}\lb_{acd}^{j_{ac} j_{cd}(-q_{ad}j_{ad})}(w)\)\)}
 \end{split}\label{m3}
\end{equation}
{}From this one can observe 
that the structure constant is independent of $\theta$ as the structure constants of the Fukaya's
$A_\infty$-category (Definition \ref{defn:fukaya}) are. 

Let us further consider 
\begin{equation}
 \ti{m}^B_3(e_{ab}^0,e_{bc}^0,e_{cd}^0)
\end{equation}
in the following situation 
$a=(\(\bps 1 & 1\\ 0 & 1\eps\), (\alpha_a,0))$, 
$b=(\(\bps 1 & 2\\ 0 & 1\eps\), (\alpha_b,0))$, 
$c=(\(\bps 1 & 0\\ 0 & 1\eps\), (0,0))$ and 
$d=(\(\bps 1 & -1\\ 0 & 1\eps\), (0,0))$. 
We have 
$g_{ab}=\(\bps 1 & 1\\ 0 & 1\eps\)$, 
$g_{bc}=\(\bps 1 & -2\\ 0 & 1\eps\)$, 
$g_{cd}=\(\bps 1 & -1\\ 0 & 1\eps\)$, 
$g_{ad}=\(\bps 1 & -2\\ 0 & 1\eps\)$ and 
$g_{ac}=\(\bps 1 & -1\\ 0 & 1\eps\)$. 
The triple product (\ref{m3}) then reduces to 
\begin{equation}
 \begin{split}
& \ti{m}^B_3(e_{ab}^0,e_{bc}^0,e_{cd}^0)=
  \sum_{v\in 2\Z}
 \sum_{w\in \Z}
\exp{\(\frac{\pi i\rhoh}{2}(\lb_{abc}^{000}(v))^2\)} \\
&\hspace*{2.0cm}
\sum_{n=1}^\infty\ov{\sqrt{2\pi} 2^n n!}
H_n\(\sqrt\frac{\pi T}{2}
\lb_{abc}^{000}(v)\)
H_{n-1}\(\sqrt\frac{\pi T}{2}
\lb_{acd}^{000}(w)\)
\exp{\(\frac{-\pi i\rhoh^*}{2}
(\lb_{acd}^{000}(w))^2\)}\ ,
 \end{split} 
\end{equation}
where $\lb_{abc}^{000}(v)=-v+2\alpha_a-\alpha_b$ and 
$\lb_{acd}^{000}(w)=-2w+\alpha_a$. 
One can also rewrite this as 
\begin{equation}
 \begin{split}
 &\sum_{\substack{s=v'-\alpha_a/2,\\ t=w'+\alpha_b/2\\v',w'\in\Z}}
 \exp{(2\pi i\rhoh(2s+t)^2)}
 \sum_{n=1}^\infty\frac{H_n\(\sqrt\frac{\pi T}{2}(2s+t)\)
H_{n-1}\(\sqrt\frac{\pi T}{2}s\)}{\sqrt{2\pi} 2^n n!}
 \exp{(-2\pi i\rhoh^*(s)^2)}\\
 &=\sum_{\substack{s=v'-\alpha_a/2,\\ t=w'+\alpha_b/2\\v',w'\in\Z}}
 \exp{(2\pi i\rhoh(3s^2+4st+t^2))}
 \sum_{n=1}^\infty\frac{H_n\(\sqrt\frac{\pi T}{2}(2s+t)\)
H_{n-1}\(\sqrt\frac{\pi T}{2}s\)}{\sqrt{2\pi} 2^n n!}
 \exp{(-2\pi T(s)^2)}
 \end{split}
\end{equation}
As in the example above, one can calculate all higher Massey 
products. 
This procedure is interesting since the results have some geometric 
interpretation in the A-model side. 
In the example above, one can see that $\ti{m}^B_3$ is given by 
connecting two kind of triangles with the propagator which is 
given by some summation of the Hermite polynomials. 
The value $\rhoh(3s^2+4st+t^2))$ is then the area of the 
square used to define the $A_\infty$-structure in the Fukaya category 
in Definition \ref{defn:fukaya}. 
However, the remaining part 
does not coincide with the one in the Fukaya category. 
These two $A_\infty$-categories should be related by a homotopy, 
which is just a basis changing and is a nonlinear isomorphism in 
supermanifold description of $A_\infty$-categories
(see\cite{infty,thesis}). 
We hope to construct it explicitly and complete the full homological mirror symmetry 
in a geometric way elsewhere.

 \section{Conclusions and Discussions}
\label{sec:CD}

In this paper, we discussed a (homological) mirror symmetry on 
noncommutative two-tori. 
From a string theory viewpoint, we clarified the meaning of noncommutativity, 
especially, its relation to the D-brane stability,  
The homological mirror symmetry was then defined precisely, 
including the cyclic symmetry. 
We proved a part of it, the compatibility of product structures with the cyclicities, 
and gave a Feynman rules which calculates all the higher (Massey) products 
on $\cC^B_\theta$ 
toward the explicit geometric proof of the homological mirror symmetry. 
Thus, the proof of the homological mirror symmetry for higher products has been left for 
a future direction. The problem is to construct an explicit homotopy between 
the minimal cyclic $A_\infty$-categories. 
 
On the other hand, expect for the higher products, 
we succeeded to have a deep understanding of 
a noncommutative mirror symmetry in the two-tori case. 
Thus, the next problem may be the extension of these results 
to higher (even) dimensional tori based on 
\cite{K,KaOr, Stheta} together with results on the Morita equivalence for higher dimensional 
noncommutative tori (see \cite{Rhigh,RS,S,PS,TW}).  
Though seemingly straightforward, we can expect that it should includes 
fruitful structures, since in this case the noncommutativity 
should not be explained by the D-brane stability only. 
It should include the abelian variety case \cite{F} and might be related 
to the lagrangian foliation \cite{FNC}. 
Also, it is interesting if we could further extend and apply our arguments 
to homological mirror symmetry for torus fibrations \cite{SYZ}.

\begin{center}
\noindent{\large \textbf{Acknowledgments}}
\end{center}

I would like to thank A.~Takahashi and K.~Fukaya 
for valuable discussions and useful comments. 
This work was completed during my stay at the Department of Mathematics 
of the University of Pennsylvania. 
I am very grateful for the support I received from the staff, students and 
faculty at the University of Pennsylvania, and especially for the 
generous hospitality extended to me by Jim Stasheff. 
The author is supported by JSPS Research Fellowships for Young Scientists.

%\newpage

\end{document}